\documentclass[twocolumn,amsmath,amssymb,superscriptaddress,aps,pra]{revtex4-2}			

\usepackage{mathtools}
\usepackage{graphicx}
\usepackage{color}
\usepackage[normalem]{ ulem }

\newcommand{\N}{\mathbb{N}}
\newcommand{\Z}{\mathbb{Z}}
\newcommand{\R}{\mathbb{R}}

\newcommand{\rd}{\mathrm{d}}

\newcommand{\betaref}{\beta_{2,\textrm{ref}}}

\newcommand\norm[1]{\| #1 \|}

\newcommand\Zref{Z_{\rm ref}}
\newcommand\epsilonZ{\varepsilon_Z}
\begin{document}

\title{
Modulational instability in optical fibers with randomly-kicked normal dispersion
}


\author{G. Dujardin}
\affiliation{Univ. Lille, CNRS UMR 8524 - Laboratoire Paul Painlev\'e, Inria, F-59000 Lille, France}
\author{A. Armaroli}
\affiliation{ Univ. Lille, CNRS, UMR 8523-PhLAM-Physique des Lasers Atomes et Mol\'ecules, F-59000 Lille, France}
\author{S. Rota Nodari}
\affiliation{Institut de Math\'ematiques de Bourgogne (IMB), CNRS, UMR 5584, Universit\'e Bourgogne Franche Comt\'e, F-21000 Dijon, France}
\author{A. Mussot}
\author{A. Kudlinski}
\affiliation{ Univ. Lille, CNRS, UMR 8523-PhLAM-Physique des Lasers Atomes et Mol\'ecules, F-59000 Lille, France}
\author{S. Trillo}
\affiliation{Department of Engineering, University of Ferrara, I-44122 Ferrara, Italy}
\author{M. Conforti}
\email{matteo.conforti@univ-lille.fr}
\affiliation{ Univ. Lille, CNRS, UMR 8523-PhLAM-Physique des Lasers Atomes et Mol\'ecules, F-59000 Lille, France}
\author{S. De Bi\`evre}
\email{stephan.de-bievre@univ-lille.fr}
\affiliation{Univ. Lille, CNRS UMR 8524 - Laboratoire Paul Painlev\'e, Inria, F-59000 Lille, France}

\begin{abstract}{
We study modulational instability (MI) in optical fibers with random group velocity dispersion (GVD)  generated by sharply localized  perturbations of a normal GVD fiber that are either randomly or periodically placed along the fiber and that have random strength. This perturbation leads to the appearance of low frequency MI side lobes that grow with the strength of the perturbations, whereas they are faded by randomness in their position. If the random perturbations exhibit a finite average value, they can be compared with periodically perturbed fibers, where Arnold tongues appear. In that case, increased randomness in the strengths of the variations tends to affect the Arnold tongues less than increased randomness in their positions. 
 }
\end{abstract}

\maketitle

\section{Introduction}
The combined effect of nonlinearity and group velocity dispersion (GVD) may lead to the destabilization of the stationary states (plane or continuous waves) of a given physical system.
This phenomenon, known under the name of modulational instability (MI), consists in the exponential growth of small harmonic perturbations of a continuous wave \cite{Zakharov2009}. MI has been pioneered in the 60s in the context of fluid mechanics \cite{Benjamin1967,Zakharov1968}, electromagnetic waves \cite{Bespalov1966} as well as in plasmas \cite{H.Ichikawa1973}, and it has been observed in nonlinear fiber optics in the 80s \cite{Tai1986}.
In uniform fibers, MI arises for anomalous (negative) GVD, but it may also appear for normal GVD if polarization \cite{1970JETP}, higher order modes \cite{Stolen1974} or higher order dispersion are considered \cite{Cavalcanti1991}.
A different kind of MI related to a parametric resonance mechanism emerges when the dispersion or the nonlinearity of the fiber are periodically modulated
\cite{Smith1996,Droques2012,Armaroli2012,Mussot2018}. Many studies were published to  address generalizations to high-order dispersion \cite{Droques2013a,Armaroli2014}, birefringence \cite{Armaroli2013}, fiber cavities \cite{Conforti2014,Conforti2016a,Copie2016,Copie2017,Copie2017a,Copie2017b,Bessin2019}, and  the nonlinear stage of MI \cite{Conforti2016}.

The effect of a random variation of GVD on MI has been studied extensively \cite{Abdullaev1996,Abdullaev1997,Abdullaev1999,Garnier2000,Chertkov2001} for the particular case where the GVD is perturbed by a Gaussian white noise, which is explicitly solvable. Under these conditions a deformation of the conventional MI gain profile due to the random perturbation was found when the unperturbed fiber has an anomalous dispersion. In the case of normal dispersion, the generation of MI sidebands as the result of the random perturbation was reported as well. White noise however, which implies arbitrarily high variations of GVD  on arbitrarily small scales, constitutes an idealization that does not always provide a relevant modeling of the randomness that may occur in physical GVD fibers~\cite{Kampen2007}.
 An attempt to consider a GVD perturbed by a Gaussian noise with a finite correlation length was reported in \cite{Karlsson1998}, but the analysis was not conclusive, and the problem was solved with numerical simulations only \cite{Farahmand2004}.

The question then arises for which type of random GVD processes MI occurs and how the characteristics of such instabilities depend on the statistical properties of the process. In such generality, the question seems however out of reach and it is consequently of interest to study the problem in a class of random fibers that is both experimentally accessible and theoretically tractable. Our focus here will therefore be on  homogeneous fibers with a normal GVD perturbed by a set of random ``kicks''. More specifically, the fibers we consider, described in more detail in Section~\ref{s:MI_random}, have a GVD given by
\begin{equation}\label{eq:beta2approximate} 
  \beta_2(z) = \beta_{2,\textrm{ref}} +  \Delta\beta_2
\sum_{n\in\Z} \lambda_n \delta \left(\frac{z-Z_n}{\Zref}\right).
\end{equation} 
In our previous work \citep{RNCDKMTDB}, we investigated such fibers with a periodic modulation of the fiber dispersion induced by a  Dirac comb, in which the $Z_n=n\Zref$  are periodically placed along the fiber and the strengths $\lambda_n$  of the delta functions are all equal. Such GVD  was shown to be well approximated in experiments by a periodic series of short gaussian-like pulses. In this work,  the $Z_n$ are chosen to be random points along the fiber, and the $\lambda_n$ are independent and identically distributed centered random variables and $\Zref$ is a reference length. We will limit our analysis to perturbations of  fibers with a normal GVD 
 because in that case the unperturbed fibers show no MI, and consequently any MI observed in the randomly perturbed fibers is entirely due to the randomness.  

The perturbations of the GVD considered in Eq.~\eqref{eq:beta2approximate} result in a non-autonomous nonlinear Schr\"odinger equation (NLSE) (see Eq.~\eqref{eq:randNLS}) determining the evolution of the wave profile as a function of the longitudinal coordinate $z$ along the fiber in which the perturbation can be interpreted as a succession of ``kicks'' taking place at points $z=Z_n$. These systems therefore bear an analogy to the paradigmatic problem of the kicked rotor in classical and quantum mechanics~\cite{Chirikov1979,QP,casati_79}, which is why  refer to them as randomly GVD kicked fibers.
 
We will show that MI occurs in such fibers through a mechanism that is familiar from Anderson localization theory \cite{Anderson1958,BoLa}. The MI gain can indeed be computed in terms of a random product of transfer matrices in the same way as the localization length of the stationary eigenfunctions of the random Schr\"odinger operator in the Anderson model.   We are then able to analyze how the properties of this gain depend on the features of the random process $\beta_2(z)$ and on the frequency of the harmonic perturbation of the continuous wave. 

The rest of the paper is organized as follows. In Section~\ref{s:MI_random} we precisely describe the randomly kicked fibers under study, then we derive a general expression for the mean MI gain in Section~\ref{s:meangain}. 
In Section~\ref{s:defoc_lambda0}, we develop a perturbative estimation of the mean MI gain in the case where the random perturbations of the GVD vanish on average  ($\overline{\lambda_n}=0$) and compare it to numerical results. We establish in this manner the existence of  MI at low frequencies of which we characterize the properties. We finally compare the  MI gain with the gain computed from the solution of the nonlinear Schr\"odinger equation and observe a good correspondence. 
In Section~\ref{s:defoclambdaplus} we consider MI  in a randomly kicked homogeneous fiber of normal GVD in which the random perturbation does not vanish on average  ($\overline{\lambda_n}\neq0$). The situation is very different since, depending on the nature of the point process $Z_n$ determining the positions of the kicks along the fiber, there may or there may not be a remnant of Arnold tongues, a signature of MI in periodic fibers. In Section V we show how GVD kicked fibers can approximate fibers with a white noise GVD. Conclusions are drawn in Section ~\ref{s:conclusions}.


\begin{figure*}
    \begin{center}
    \begin{tabular}{ccc}
      \includegraphics[width=0.33 \textwidth,keepaspectratio]{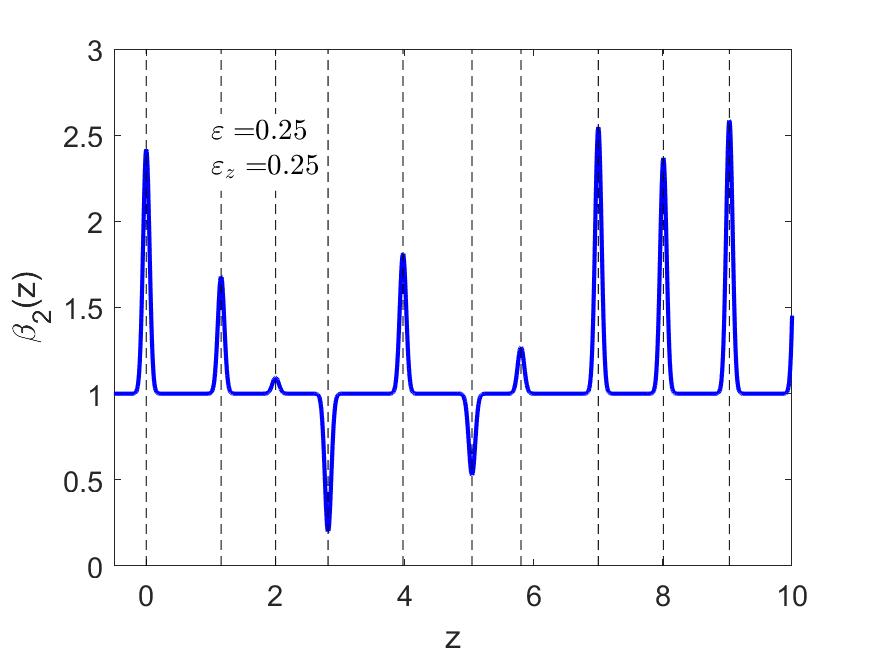}
      \includegraphics[width=0.33 \textwidth,keepaspectratio]{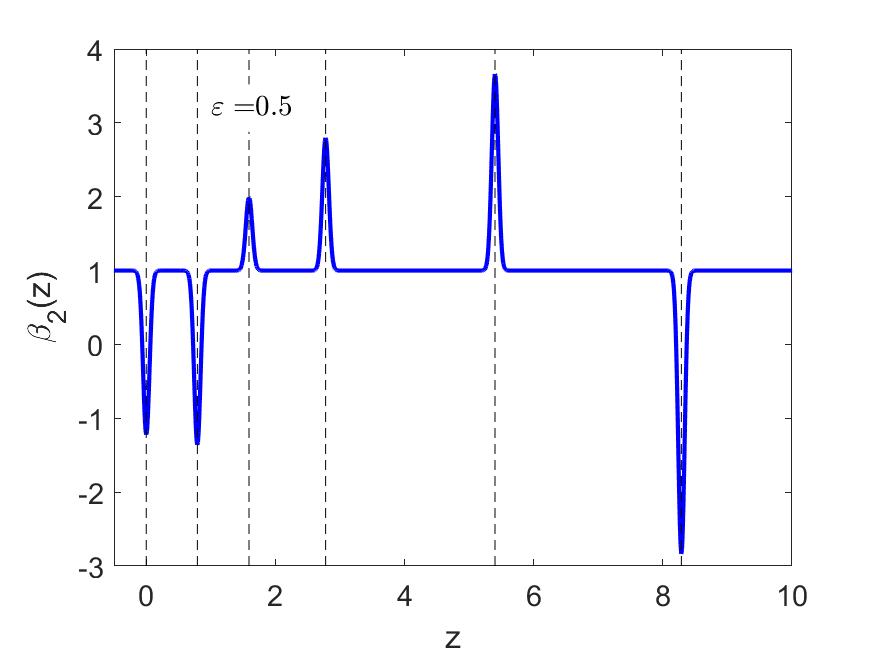}
       \includegraphics[width=0.33 \textwidth,keepaspectratio]{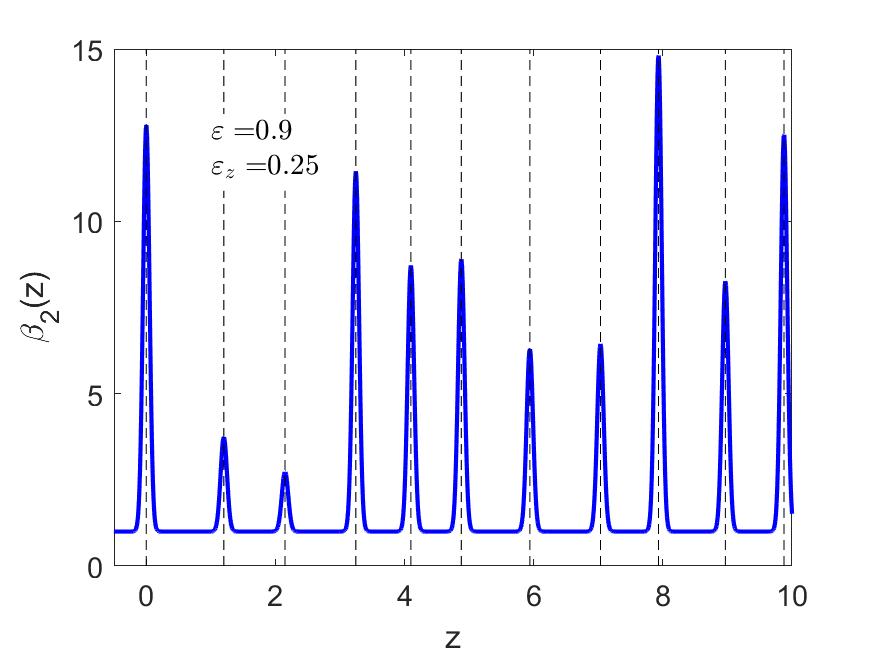}
    \end{tabular}
  \end{center}
  \caption{Realization of a simple random walk fiber (left panel) and a Poisson fiber (central panel) with zero-mean kick strength $\bar\lambda=0$, and of a simple random walk fiber with $\bar\lambda=1$ (right panel). The kicks are modeled with Gaussian pulses $\delta(z)=1/\sqrt{2\pi w^2}\,\exp(-z^2/\,2w^2)$ of width $w=\Zref/20$. The position of the kicks is indicated by the vertical dotted lines;  
 \label{fig:sketch}
  $\lambda_n$  are uniform random variables in $[-1,1]$. }
\end{figure*}

\section{Modulational instability in randomly kicked fibers}\label{s:MI_random}
We consider the NLSE
\begin{equation}
\label{eq:randNLS}
  i\partial_z u -\frac{1}{2}\beta_2(z)
\partial_t^2 u + \gamma |u|^2 u
= 0,
\end{equation}
where $\gamma>0$ is the fiber nonlinear coefficient, and $\beta_2(z)$ its GVD. We are interested in the modulational (in)stability of the stationary solution  $u_0(z)=\sqrt P\exp(iP\gamma z)$ of Eq.~\eqref{eq:randNLS}. 
We consider a perturbation of $u_0(z)$ in the form $u(z,t)=[v(z,t)+1]u_0(z)$,  where the perturbation $v(z,t)$ satisfies $|v|\ll 1$. Writing $v=q+ip$, with $q$ and $p$ real functions, inserting this expression into Eq.~\eqref{eq:randNLS}, and retaining only the linear terms, we obtain a linear system for $q$ and $p$. Writing $x(z,t)=(q(z,t), p(z,t))$ and $\hat x(z,\omega)=\frac{1}{\sqrt{2\pi}}\int x(z,t)e^{-i\omega t}\,\rd t$ one finds
\begin{equation} \label{nlsefiberlinfouriersys}
\partial_z\hat x(z,\omega)=
\begin{pmatrix}
0&-\frac{\beta_2(z)}{2}\omega^2\\ \frac{\beta_2(z)}{2}\omega^2+2\gamma P&0
\end{pmatrix} \hat x(z,\omega).
\end{equation}
{\color{blue} }\\
 Note that this is, for each $\omega$, a non-autonomous linear Hamiltonian dynamical system in a two-dimensional phase plane with canonical coordinates $(\hat q, \hat p)$. We wish to study the (in)stability of its fixed point at the origin $\hat q=0=\hat p$, as a function of the frequency $\omega$ and of the properties of $\beta_2(z)$. Since for general $\beta_2(z)$, its explicit solution cannot be computed analytically, this is not straightforward. We will concentrate on random fiber profiles, as detailed below. 
 
Note that if $\gamma=0$, Eq.~\eqref{nlsefiberlinfouriersys} is reminiscent of a harmonic oscillator with random frequency $k=\frac{\beta_2(z)}{2}\omega^2$ (also called multiplicative noise)
for which a vast literature exists \cite{VanKampen1976,Bobryk2002,Gitterman2005,Mallick2006,Kampen2007,Poulin2008}. The focus of the present work is to study a ``generalised''  random oscillator including the term $2\gamma P$) which accounts for the nonlinear effects, where the random frequency is modeled by the non-stationary stochastic processes described by Eq.~\eqref{eq:beta2approximate}. The study of stationary colored noise, where classic perturbative techniques apply \cite{Kampen2007}, will be the subject of a future work \cite{newpaper}.

 The modulational instability of the fiber is expressed in terms of the sample MI gain $G(\omega)$, defined as follows:
 \begin{equation}\label{eq:generalgain}
 G(\omega)=\lim_{z\to+\infty}\frac1{z}\ln\|\hat x(z,\omega)\|,
 \end{equation}
 where $\norm{\cdot}$ designates the euclidean norm. Here sample stands for a single realization of the random perturbation. When $G(\omega)>0$, this indicates $\|\hat x(z,\omega)\|\simeq \exp(G(\omega)z)$, meaning that the stationary solution is unstable for perturbations with frequency $\omega$. One is interested in establishing for which $\omega$, if any, this occurs, and how large $G(\omega)$ is in that case.

\emph{Dispersion-kicked fibers} are characterized by the expression of $\beta_2(z)$ given in Eq.~(\ref{eq:beta2approximate}). 
Here $\delta$ is a sharply peaked 
positive function satisfying $\int_\R\delta(z)\rd z=1$; in our theoretical analysis below, we will take $\delta$ to be a Dirac delta function; 
$\lambda_n$ are independent, identically distributed real random variables with mean $\overline \lambda\geq0$,
and $\Zref>0$ is a characteristic length
associated to the random sequence of points $Z_n$. We will write
\begin{equation}\label{eq:lambdan}
\lambda_n=\overline\lambda +\varepsilon\delta\lambda_n, 
\end{equation}
with $\overline{\delta\lambda}_n=0$ and $\varepsilon>0$ a dimensionless parameter. 
We think of this as a fiber with irregularities in its diameter of random area, 
 giving rise to effective dispersion kicks $|\lambda_n|\Delta\beta_2 \Zref$, placed at the points $Z_n$ along the fiber. 
 These fibers can be physically fabricated by means of the state-of-the-art fiber-drawing techniques. Indeed, some examples of uniformly dispersion-kicked fibers has been reported in \cite{RNCDKMTDB} with a period $\Zref=10$ m and a kick width $w=0.14$ m and relative large kick strength $\max |\beta_2(z)|/\betaref\approx35$.



To compute the MI gain in such fibers, we proceed as follows. We first note that, due to the presence of the delta functions, the left and right limits $\hat x_n^\pm(\omega)=(\hat{q}(Z_n^\pm,\omega),\hat{p}(Z_n^\pm,\omega))$ of the solution $\hat x(z,\omega)$ at $Z_n$ are different and are related by $\hat x_n^+(\omega)=K_n\hat x_n^-(\omega)$, where the random matrix $K_n$ is defined as
\begin{equation}\label{eq:kickmatrix}
    K_n=\left(
    \begin{matrix}
      \cos( \Delta\beta_2\Zref\lambda_n \frac{\omega^2}{2})
& -\sin( \Delta\beta_2\Zref\lambda_n \frac{\omega^2}{2}) \\
      \sin(\Delta\beta_2\Zref\lambda_n \frac{\omega^2}{2})
& \cos( \Delta\beta_2\Zref\lambda_n \frac{\omega^2}{2})
    \end{matrix}
\right).
\end{equation}
On the other hand, for $Z_n<z<Z_{n+1}$, \eqref{nlsefiberlinfouriersys} is autonomous and straightforwardly solved; the solution $\hat x(z,\omega)$ is smooth in this range. One finds $\hat x_{n+1}^-=L_n\hat x_n^+$, where now
\begin{equation}
  L_n=\left(
    \begin{matrix}
      \cos(k \Delta Z_n) & -\mu\sin(k\Delta Z_n) \\
      \mu^{-1}\sin(k\Delta Z_n) & \cos(k\Delta Z_n)
    \end{matrix}
\right),\label{eq:transportmatrix}
\end{equation}
with
\begin{equation}\label{eq:k2omega}
  k^2=\frac{\betaref}{2} \omega^2 \left(\frac{\betaref}{2}\omega^2 + 2\gamma P\right), \; \mu=\frac{\frac{\betaref}{2} \omega^2}{k}.
\end{equation}
Note that $k$ is real for all $\omega$ when $\betaref>0$ (normal dispersion or defocusing NLSE), and that it is imaginary for small $\omega$ when $\betaref<0$ (anomaluous dispersion or focusing NLSE). 
To sum up, we can now describe the evolution of this system between $z=Z_n^-$ and
$z=Z_{n+1}^-$ as follows. For all $\omega\in\R$ and $n\in\N$
\begin{equation} \label{eq:transfermatrix}
\hat x_{n+1}^-=  
L_nK_n \hat x_n^-=\Phi_n \hat x_n^-.
\end{equation}
Considering an initial condition $\hat x_0^-=(\hat{q}(Z_0^-,\omega), \hat{p}(Z_0^-,\omega))\in\R^2$ with $\norm{\hat x^-_0}=1$, one then finds
\begin{equation}
 G(\omega)= \frac1{\Zref}\lim_{n\to{+\infty}}\frac1n \ln \norm{\Phi_n \Phi_{n-1} \dots \Phi_1\hat x^-_0 }. \label{eq:kickgain}
\end{equation}

Note that
$$
0\leq \mu \leq 1, \; \mu\simeq 1-\frac{2\gamma P}{\betaref\omega^2}+ O(\omega^{-4}), \, \lim_{\omega\to+\infty }\mu =1.
$$
Hence it follows that for large $\omega$, both $K_n$ and $L_n$ are rotation matrices and consequently their product $\Phi_n$ is also a rotation matrix. Consequently, for large $\omega$, the MI gain tends to zero.

We finally describe the models we will consider for the random positions  $Z_n$ of the delta-functions, and for their strengths $\lambda_n$. For the $Z_n$, we set
  \begin{equation}\label{eq:RW}
  Z_0=0,\;
        \forall n\in\N,\ Z_{n+1}=Z_{n}+\Zref j_n,
 \end{equation}
where $\Zref>0$ and $j_n$ is a sequence of independent and identically distributed positive
random variables with values in $\R^+$ and with probability density $\rho(j_n)$.
We assume that 
\begin{equation}\label{eq:RWfiber}
\langle j_n\rangle =\int_0^{+\infty} j\rho(j)\rd j=1,\; \\
\end{equation}
so that $\langle \Delta Z_n\rangle = \langle(Z_{n+1}-Z_n) \rangle= \Zref>0$.
Hence $Z_{n+1}=Z_n+ \Delta Z_n$ is a random walk with drift.  We will principally consider two cases. First, introducing a new parameter $\epsilonZ>0$, we consider
\begin{equation}\label{eq:RWstepdistribution}
  j_n = \left(1 + \epsilonZ \delta\!j_n\right),
\end{equation}
where now $\delta\!j_n$ is a sequence of independent, identically and uniformly distributed random variables in $(-1,1)$ and $\epsilonZ\in [0,1]$. In this random walk model the mean position of $Z_n$ is $n\Zref$ and  their variance
$$
\langle \left(Z_n-n\Zref\right)^2\rangle=n\epsilonZ^2\Zref^2\langle(\delta\! j)^2\rangle,
$$
grows with $n$. The increments are independent and identically distributed and their variance is given by
 $$
\langle (\Delta Z_n-\Zref)^2 \rangle= \Zref^2 \epsilonZ^2 \langle(\delta\!j)^2 \rangle.
$$
We will refer to this as the simple random walk model.

Second, we will consider the Poisson model where the $\Delta Z_n=\Zref j_n$ are independent and identically distributed with exponential density
\begin{equation}
\rho(j)=\exp(-j).\label{eq:RWPoissonstep}
\end{equation}
The $Z_n$ are now the arrival times of a Poisson process with parameter $\Zref^{-1}$. We will refer to this as the Poisson fiber. 

When all $\lambda_n=0$ the above fibers are homogeneous.  
We will concentrate here on the defocusing regime, in which no modulational instability occurs when $\lambda_n=0$. 

Some examples of realizations of the random walk and of the Poisson fibers are reported in Fig. \ref{fig:sketch}, where the kicks are modeled by sharp Gaussian functions. Here and in all numerical examples we take $\Delta\beta_2=\betaref=\gamma=P=\Zref=1$.

In Section~\ref{s:meangain} and~\ref{s:defoc_lambda0} we consider random kick strengths $\lambda_n$ that vanish on average. Our focus is therefore on the question: what kind of random inhomogeneities  of the GVD can produce MI in an otherwise modulationally stable homogeneous fiber?  

In Section~\ref{s:defoclambdaplus} we then briefly discuss the case $\overline\lambda>0$: in that situation, MI occurs in the form of Arnold tongues in the limiting case when $\epsilonZ=0$, 
resulting in a periodic GVD. We will investigate the stability of these Arnold tongues under the random perturbations of the $Z_n$ and/or $\lambda_n$ which occur when $\epsilonZ\not=0$ and/or $\varepsilon\not=0$.

\section{The mean MI gain of a randomly kicked fiber}
\label{s:meangain}


 Since the random walk process $Z_n$ in Eq.~\eqref{eq:RWfiber} has independent and identically distributed increments the Furstenberg theorem~\cite{Fur} (see~\cite{BoLa} for a textbook treatment) asserts that the limit in Eq.~(\ref{eq:kickgain}) exists for almost every realization of the fiber, that is, for amost every choice of the $\lambda_n$ and the $j_n$; it is in addition strictly positive and independent of the realization.
  As a result, in such random fibers, there is always MI at all values of $\omega$. However, it is notoriously difficult to obtain analytical expressions for $G(\omega)$ as function of the model parameters, and hence to assess the strength of the sample MI gain $G(\omega)$. We will therefore follow \cite{Garnier2000} and introduce a suitable mean MI gain based on moments of Eq.~\eqref{eq:transfermatrix} which we will refer to as $G_2(\omega)$.

We recall that no useful information can be obtained from the evolution of first-order moments of Eq.~\eqref{eq:transfermatrix}. The second-order moments have instead to be computed. Let
\begin{equation*}
\Phi_n\equiv L_nK_n=
  \left(
  \begin{matrix}
  a_n & b_n\\
  c_n & d_n
  \end{matrix}
  \right).
\end{equation*}
It is then straightforward to check that 
for all $n\in\N$,
\begin{equation}\label{eq:meanevolution}
  X_{n+1}:= \left(
   \begin{matrix}
        \hat q(Z_{n+1}^-)^2\\
       \hat q(Z_{n+1}^-) \hat p(Z_{n+1}^-)\\
       \hat p(Z_{n+1}^-)^2
    \end{matrix}
  \right)
=M_n
   \left(
    \begin{matrix}
        \hat q(Z_{n}^-)^2\\
      \hat q(Z_{n}^-) \hat p(Z_{n}^-)\\
       \hat p(Z_{n}^-)^2
    \end{matrix}
  \right),
\end{equation}
where
\begin{equation}
 M_n=
  \begin{pmatrix}
    a_n^2 & 2 a_n b_n & b_n^2\\
    a_n c_n& a_n d_n + b_n c_n & b_n d_n\\
  c_n^2 & 2 c_n d_n & d_n^2
  \end{pmatrix}.
\end{equation} 
Since $\|\hat x_n^-\|^2\le\|X_n\|_1\le\frac{3}{2}\|\hat x_n^-\|^2 $ it readily follows that 
\begin{equation*}
G(\omega)=\frac1{2\Zref}\lim_{n\to{+\infty}} \frac1n \ln \|X_n\|_1,
\end{equation*}
where now $\|(x,y,z)\|_1=|x|+|y|+|z|$. From Jensen's inequality, it follows that 
\begin{equation}\label{eq:GGbarineq}
\begin{aligned}
G(\omega)&=\frac1{2\Zref}\lim_{n\to{+\infty}}  \frac1n\overline{\ln \|X_n\|_1}\\ 
&\leq \frac1{2\Zref}\lim_{n\to{+\infty}}  \frac1n\ln \overline{\|X_n\|_1} = G_2(\omega).
\end{aligned}
\end{equation}
We will refer to $G_2(\omega)$ as the mean MI gain. It is worth to point out that this corresponds to one half the growth rate of the average power of the perturbations.

As we now show, $G_2(\omega)$ is larger than the sample MI gain $G(\omega)$ itself but can be more readily computed. For that purpose, first note that, since $|\hat q(Z_{n}^-) \hat p(Z_{n}^-)|\leq \frac12((\hat q(Z_{n}^-))^2+ (\hat p(Z_{n}^-))^2)$, and since
$\left|\overline{\hat q(Z_{n}^-) \hat p(Z_{n}^-)}\right|\leq \overline{|\hat q(Z_{n}^-) \hat p(Z_{n}^-)|}$, we have 
$$
\|\overline{X_n}\|_1\leq \overline{\|X_n\|_1}\leq \frac32 \|\overline{X_n}\|_1.
$$ 
Since the  $M_n$ are identically distributed and mutually independent,
  one has moreover  $\overline{X_n}=\overline{M_n}\ \overline{X_{n-1}}=\overline M^n X_0$, where
\begin{equation}\label{eq:meanM}
\overline M= 
\left(
  \begin{matrix}
    \overline{a_n^2} & 2 \overline{a_n b_n} & \overline{b_n^2}\\
    \overline{a_n c_n}& \overline{a_n d_n} + \overline{b_n c_n} & \overline{b_n d_n} \\
 \overline{c_n^2}  & 2 \overline{c_n d_n} & \overline{d_n^2} 
  \end{matrix}
\right)
\end{equation}
is independent of $n$. Hence
\begin{align}\label{eq:meangain}
\nonumber G_2(\omega)& =\frac1{2\Zref}\lim_{n\to{+\infty}} \frac1n\ln \|\overline{X_n}\|_1\\
                   & =\frac1{2\Zref}\lim_{n\to{+\infty}} \frac1n\ln \|\overline{M}^n X_0\|_1=\frac{1}{2\Zref}\ln|x|,
\end{align}
where $x$ is the eigenvalue of $\overline M$ with largest modulus. 

 The matrix $\overline M$ depends on the parameters of the model, in particular on $\omega$, the laws of $\lambda_n, j_n$, $\varepsilon$ and $\epsilonZ$. \\
It follows that $G_2(\omega)$ can be computed from the spectrum of $\overline M$, which is easily determined numerically. The explicit analytical computation of closed formulas for its spectrum remains however complicated. 
As we show below, a perturbative treatment yields an analytic expression provided that the perturbation is not too large.

%
%

\begin{figure*}
  \begin{center}
    \begin{tabular}{cccc}
      \includegraphics[width=0.25 \textwidth,keepaspectratio]{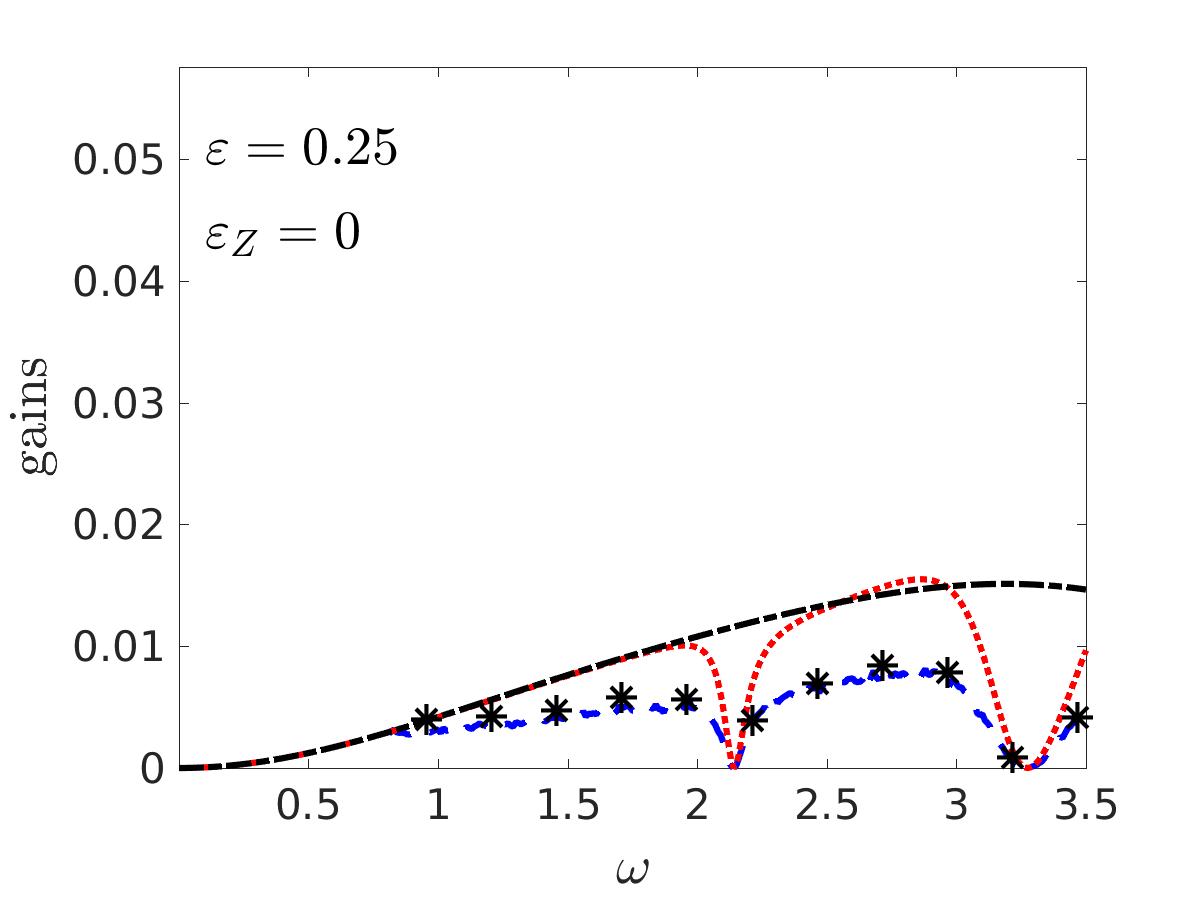}&
      \includegraphics[width=0.25 \textwidth,keepaspectratio]{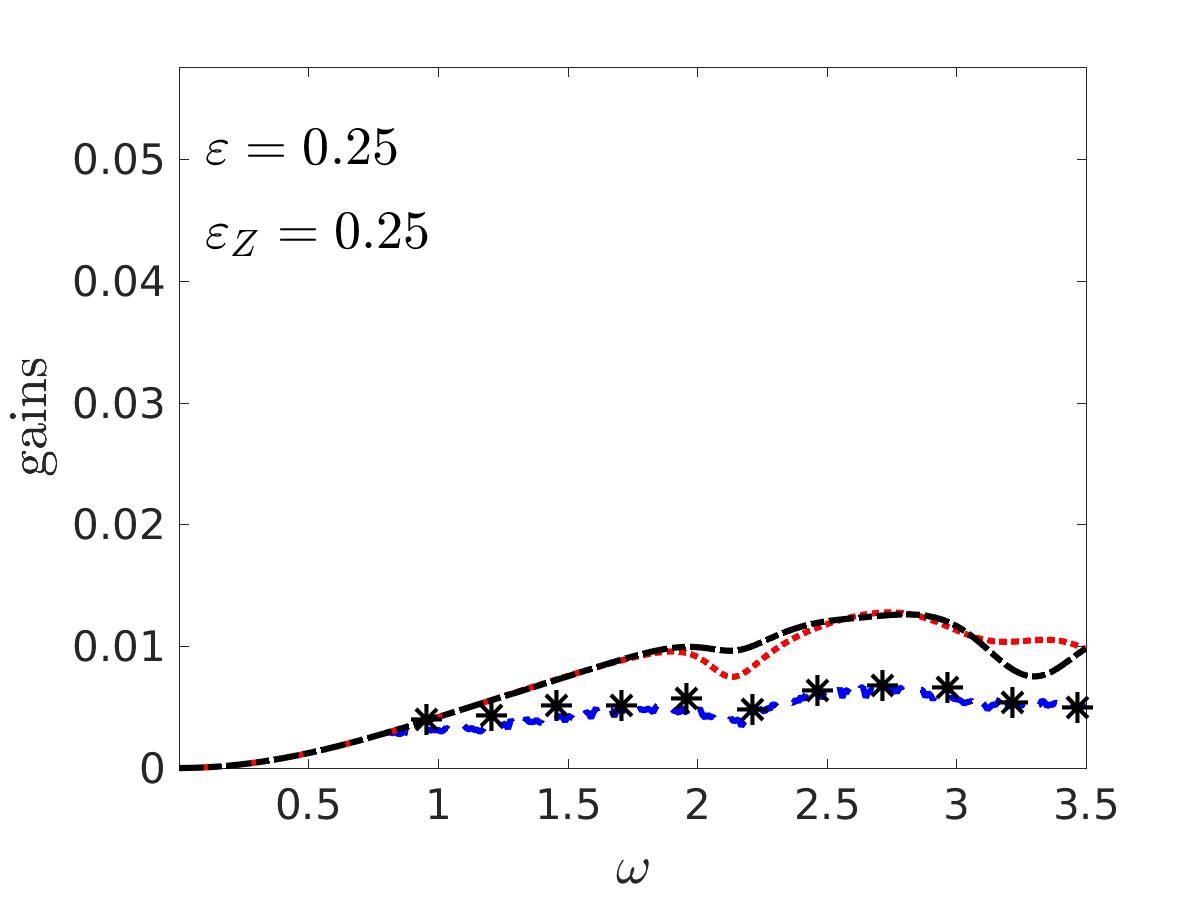}&
      \includegraphics[width=0.25 \textwidth,keepaspectratio]{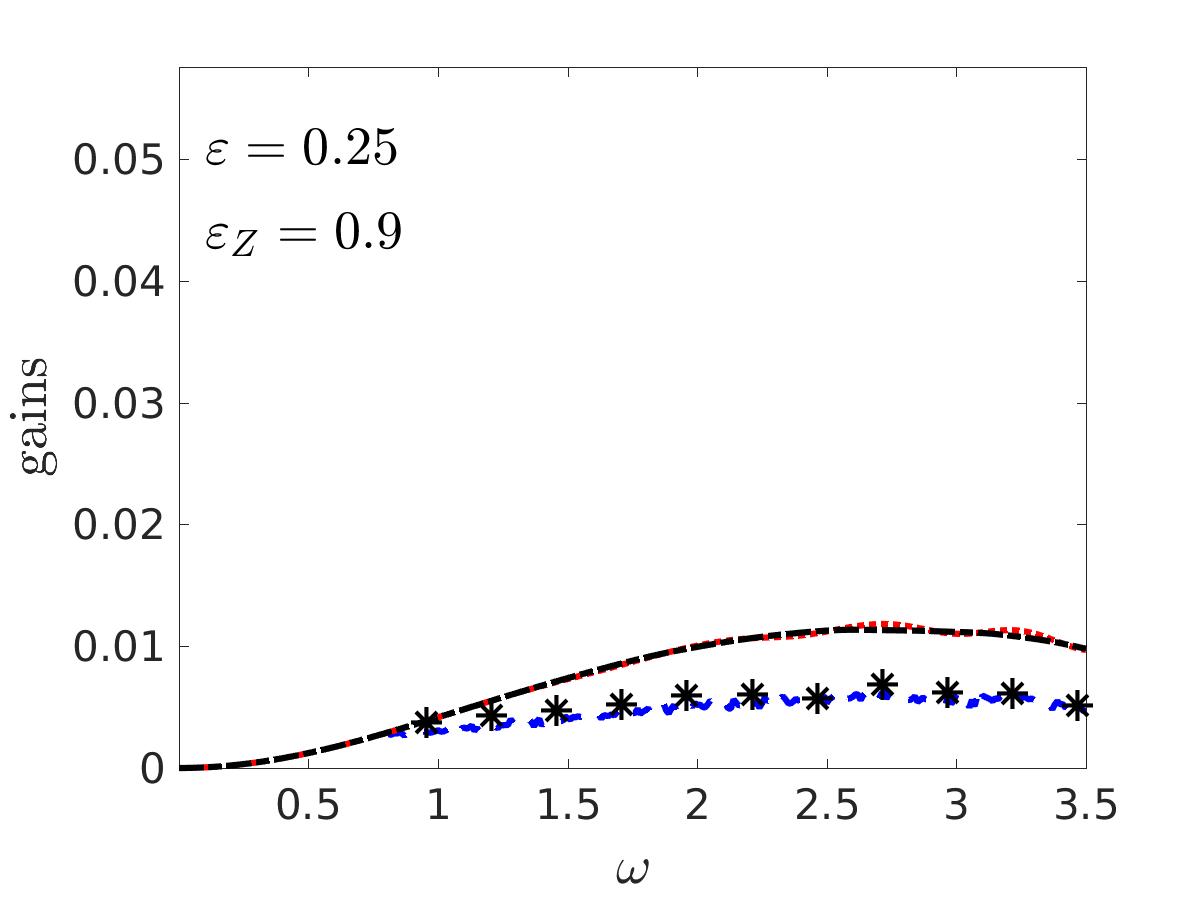} \\
      \includegraphics[width=0.25 \textwidth,keepaspectratio]{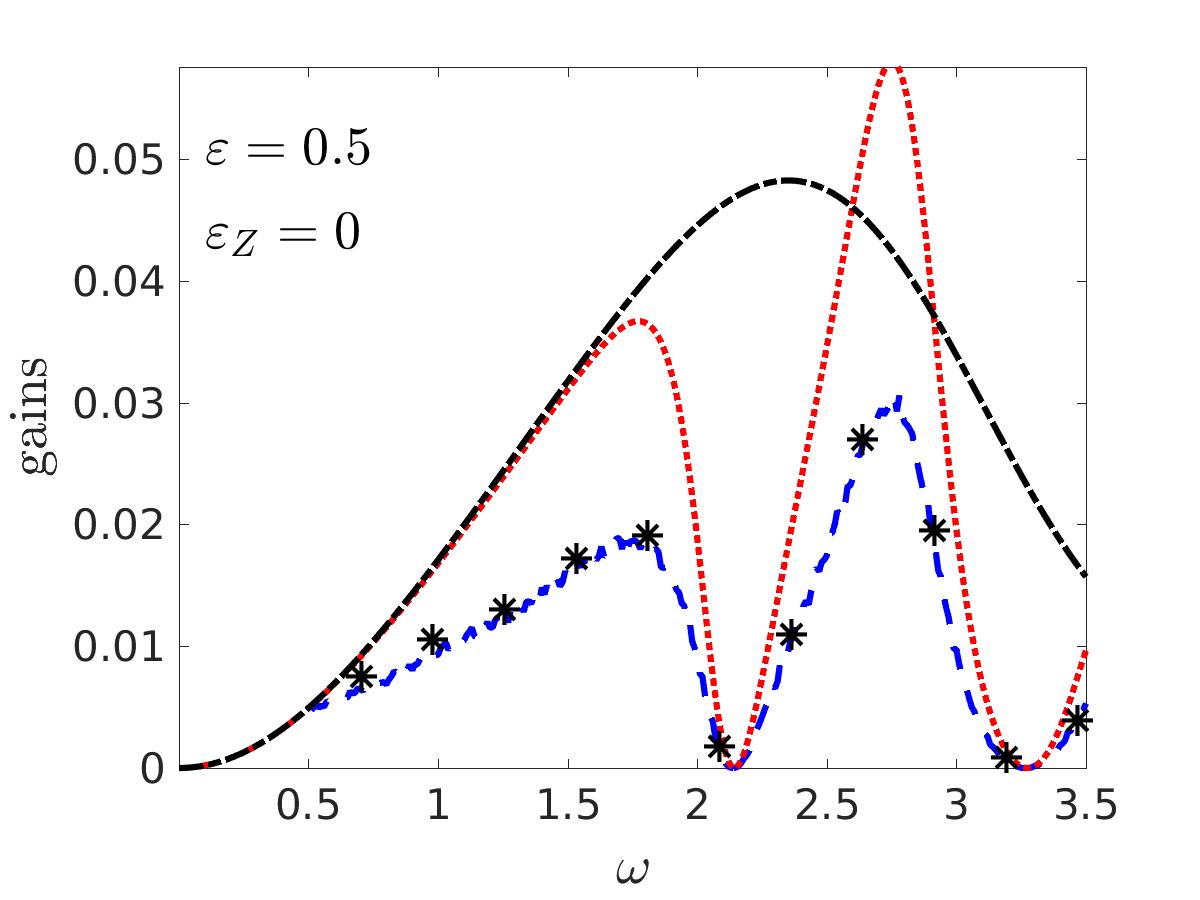}&
      \includegraphics[width=0.25 \textwidth,keepaspectratio]{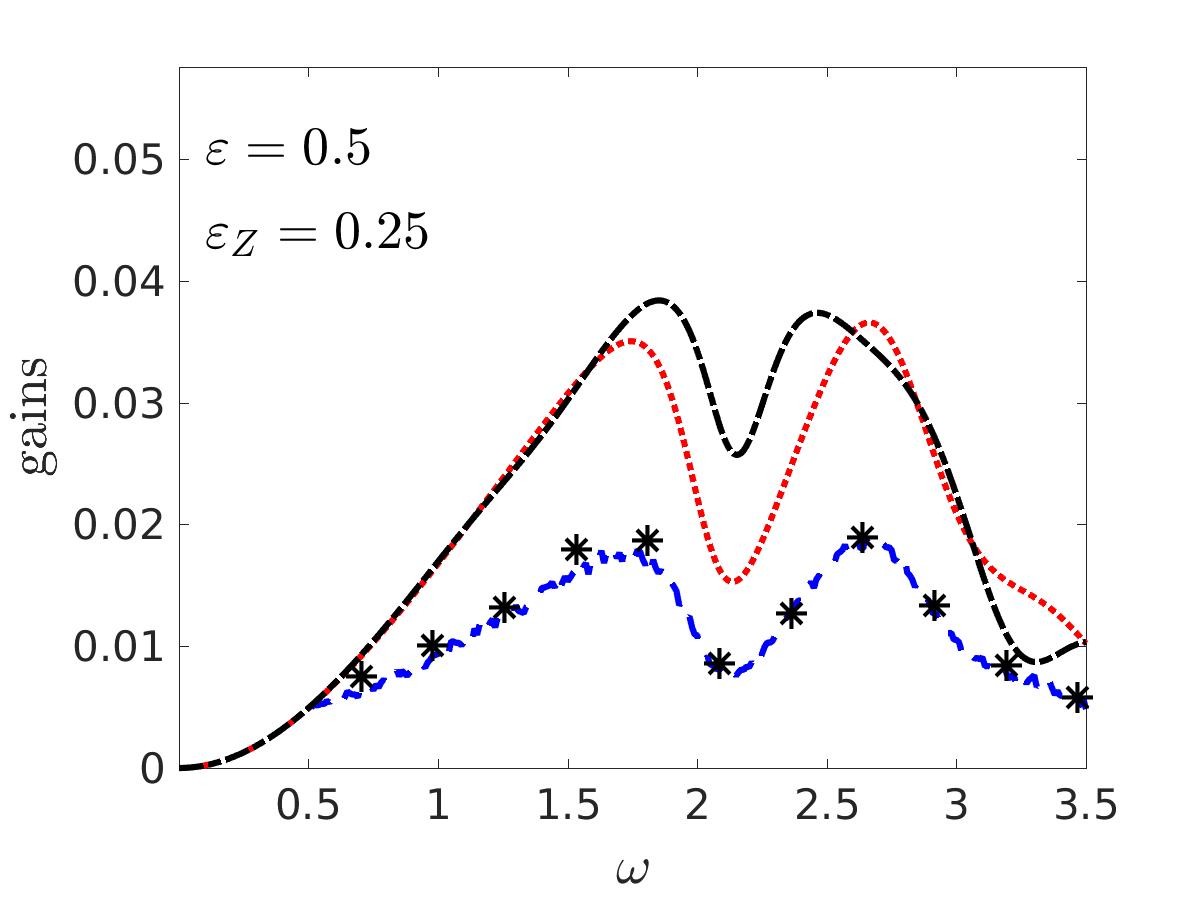}&
      \includegraphics[width=0.25 \textwidth,keepaspectratio]{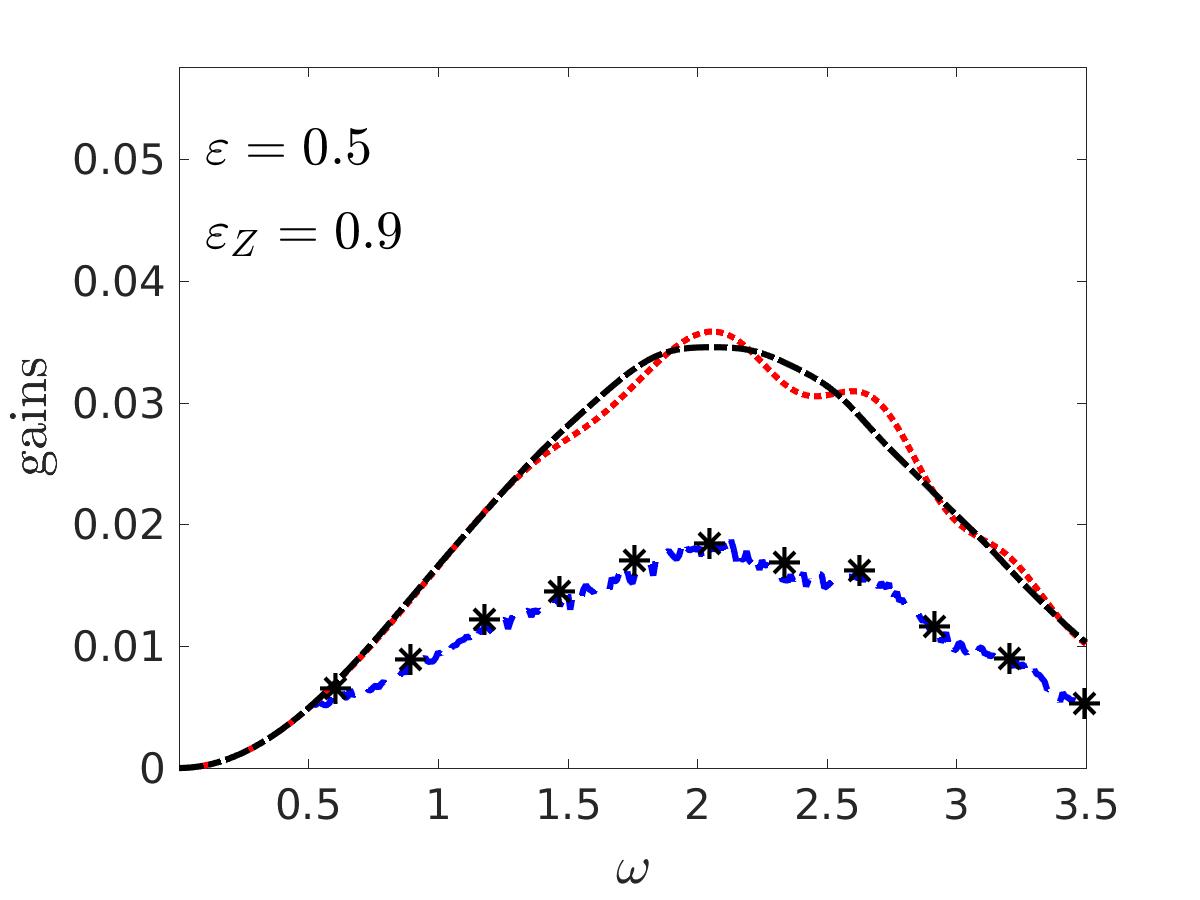}\\
    \end{tabular}
  \end{center}
  \caption{Comparison of numerically computed MI gains in 
    randomly kicked homogeneous fibers with $\overline\lambda=0$;
    $\epsilonZ$ and $\varepsilon$ are indicated in each panel.
    The random kicks are located at $Z_n$ as in Eq.~\eqref{eq:RWstepdistribution} (random walk).
    The $\delta\lambda_n$ and $\delta\! j_n$ are taken to be uniform in $[-1,1]$. 
    The numerically simulated sample MI gain $G(\omega)$ (dashed blue line),
    the mean MI gain $G_2(\omega)$ computed directly from the
    largest eigenvalue of $\overline M$ (dotted red line),
    the perturbative approximation of the mean MI gain (dash-dot black line)
    and the perturbations' growth rates (black stars) calculated from numerical solutions of the NLSE 
    are shown.
  \label{fig:lambdabar0_RW}}
\end{figure*}

Dropping the index $n$ on $\lambda_n, Z_n$, we introduce 
\begin{equation}\label{eq:theta}
\theta= \Delta\beta_2\Zref\lambda\frac{\omega^2}{2},
\end{equation}
which corresponds to the phase acquired by the perturbation $\hat v(\omega)$ at each kick, and write
\begin{equation}\label{eq:Mtheta}
M=M_+\cos^2\theta +M_0\sin(2\theta) +M_-\sin^2\theta,
\end{equation}
where
 \begin{equation*}
 M_+=
\begin{pmatrix}
{\cos^2(k\Delta Z)}&-\mu {\sin(2k\Delta Z)}& \mu^2{\sin^2(k\Delta Z)}\\
\frac{\sin(2k\Delta Z)}{2\mu}&\cos(2k\Delta Z)&-\frac {\mu\sin(2k\Delta Z)}{2}\\
\frac{\sin^2(k\Delta Z)}{\mu^2}&\frac {\sin(2k\Delta Z)}{\mu}&{\cos^2(k\Delta Z)}
\end{pmatrix}
,    
  \end{equation*}
\begin{equation*}
 M_-= 
\begin{pmatrix}
\mu^2 {\sin^2(k\Delta Z)}&
\mu{\sin(2k\Delta Z)}&
{\cos^2(k\Delta Z)}\\
-\frac{\mu\sin(2k\Delta Z)}{2}& 
-\cos(2k\Delta Z)&
\frac{\sin(2k\Delta Z)}{2\mu}\\
{\cos^2(k\Delta Z)}&
-\frac{\sin(2k\Delta Z)}{\mu}& \frac {\sin^2(k\Delta Z)}{\mu^2}
\end{pmatrix}
,    
  \end{equation*}
and
\begin{align*}
& M_0=\\ &\left(
\begin{matrix}
-\frac{\mu\sin(2k\Delta Z)}{2}&
\mu^2\sin^2(k\Delta Z)-\cos^2(k\Delta Z)&\frac{\mu \sin(2k\Delta Z)}{2}\\
\frac{\cos(2k\Delta Z)}{2}&
-(\mu+\frac{1}{\mu})\frac{\sin(2k\Delta Z)}{2}&
-\frac{\cos(2k\Delta Z)}{2}\\
\frac{\sin(2k\Delta Z)}{2\mu}&
\cos^2(k\Delta Z)-\frac{\sin^2(k\Delta Z)}{{\mu^2}}&
-\frac{\sin(2k\Delta Z)}{2\mu}
\end{matrix}
\right).
  \end{align*}
From the independence of the random variables it follows that 
\begin{equation}\label{eq:averageM}
\overline M= \overline M_+\overline{\cos^2\theta}+\overline M_0 \overline{\sin2\theta} +\overline M_-\overline{\sin^2\theta}.
\end{equation}
The expressions of the coefficients of the matrix $\overline M$ are reported in Appendix \ref{M}; the (in)stability depends on  its spectrum. 
Note that all information on the randomness in the strengths of the kicks is contained in $\theta$, whereas the randomness in the spacings between the kicks is encoded in the matrices $M_0, M_\pm$.

For a homogeneneous randomly kicked fiber with average normal GVD and  vanishing mean kicking strength ($\overline\lambda=0$) we resort to a perturbative analysis, presented below in Sec.~\ref{s:defoc_lambda0}.

\section{Zero average kick amplitude}\label{s:defoc_lambda0}
\begin{figure*}
  \begin{center}
    \begin{tabular}{ccc}
      \includegraphics[width=0.25 \textwidth,keepaspectratio]{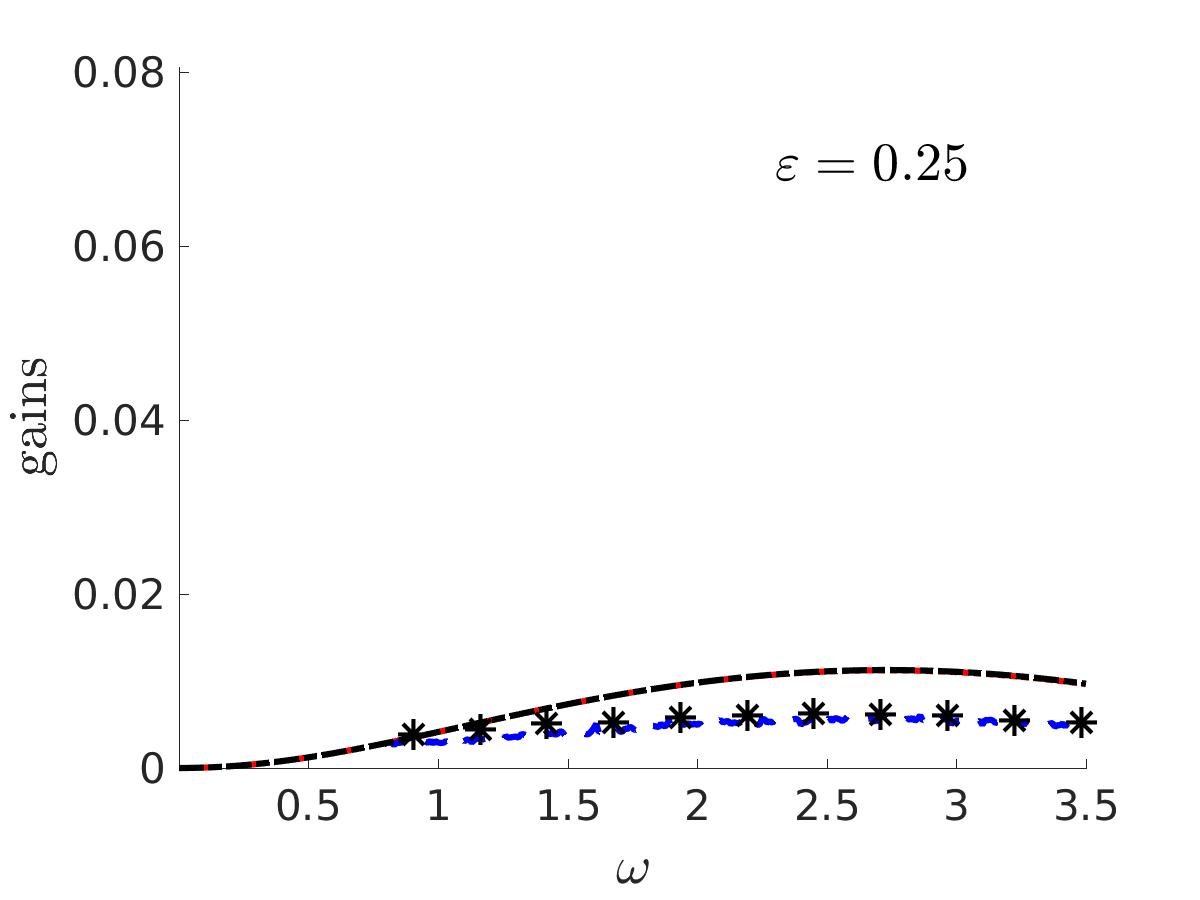}&
      \includegraphics[width=0.25 \textwidth,keepaspectratio]{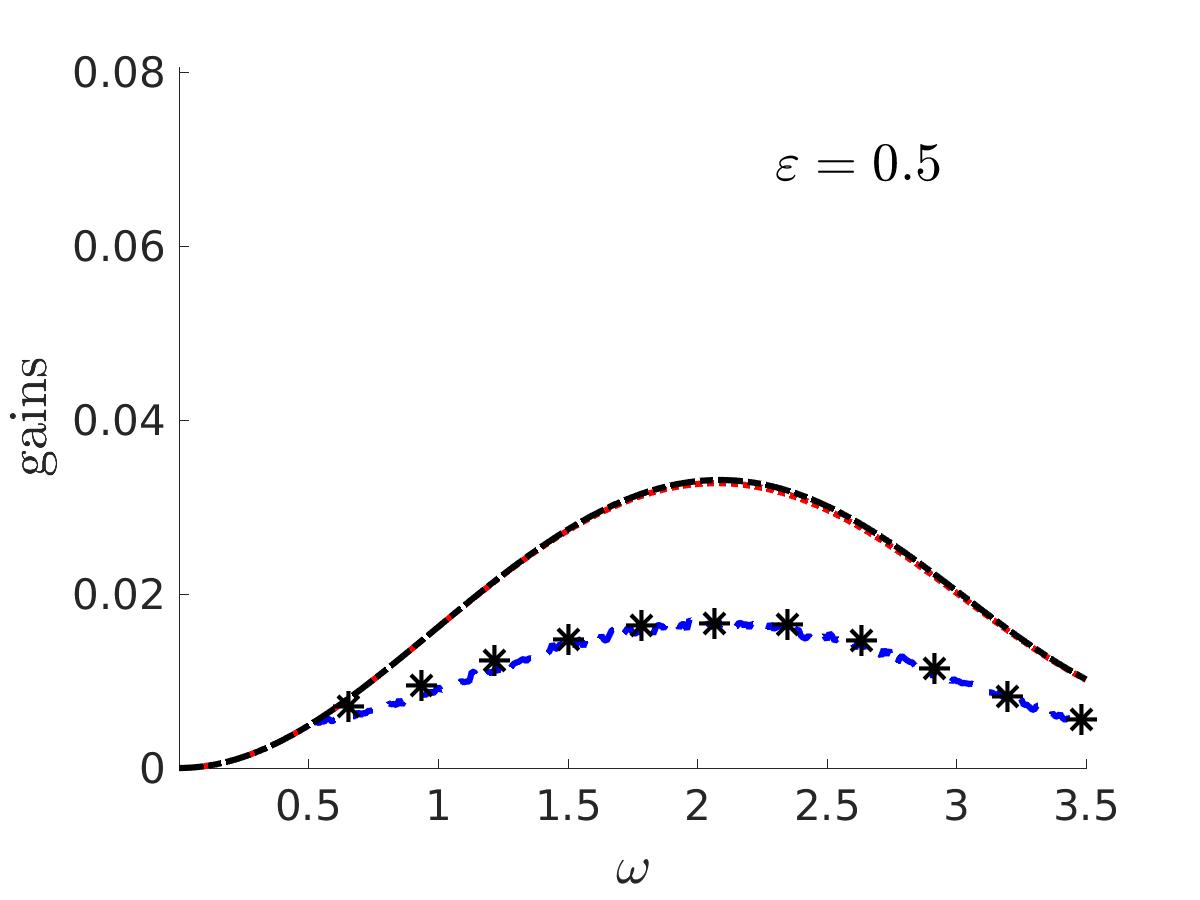}&
      \includegraphics[width=0.25 \textwidth,keepaspectratio]{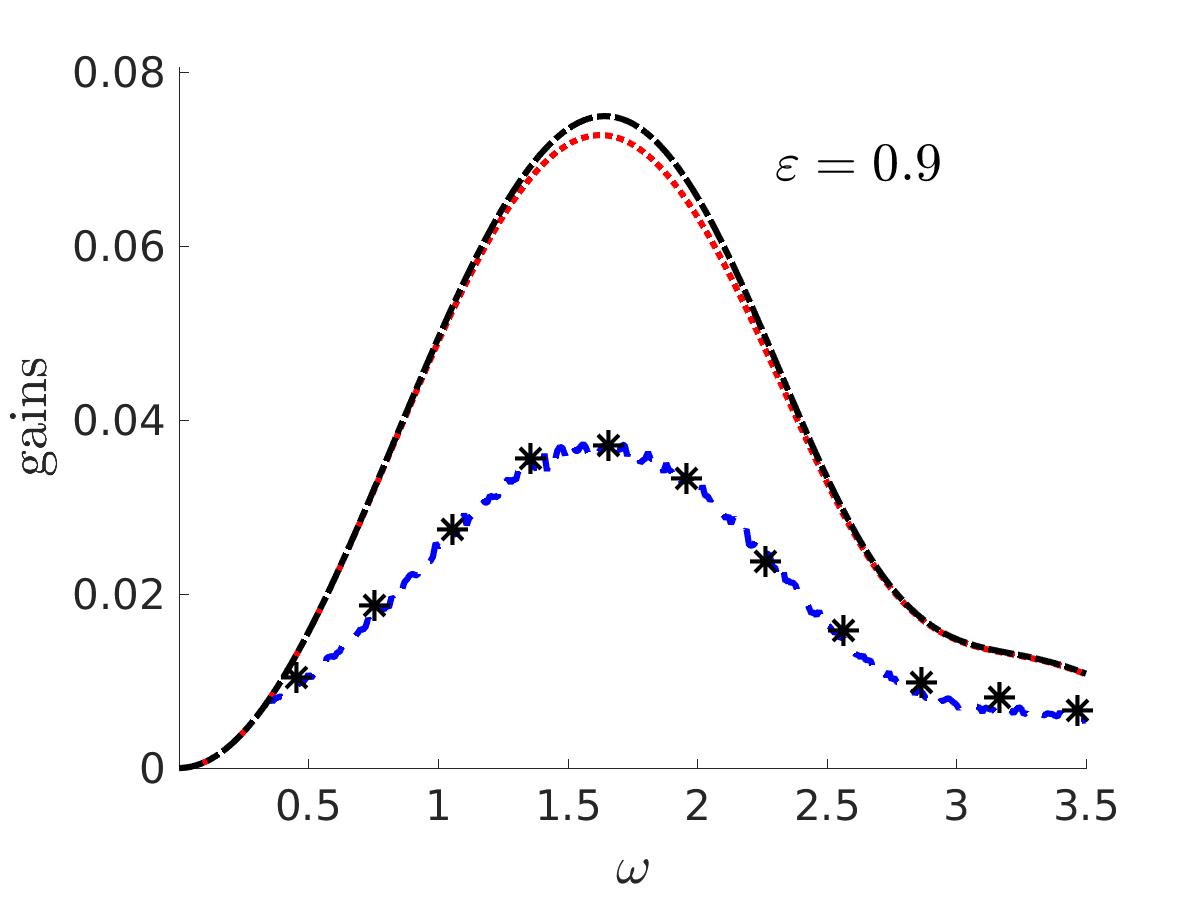}
    \end{tabular}
  \end{center}
  \caption{Same as Fig.~\ref{fig:lambdabar0_RW}, with $Z_n$ as in Eq.~\eqref{eq:RWPoissonstep} (Poisson model).
    The parameter  $\varepsilon$ is indicated in the panels.
\label{fig:lambdabar0_Poisson}}
\end{figure*}

 We consider in this section randomly kicked fibers as in Eq.~\eqref{eq:beta2approximate} with $\lambda_n$ as in Eq.~\eqref{eq:lambdan} and $\overline\lambda=0$ and with $Z_n$ a random process as in Eq.~\eqref{eq:RW}, so that the random fiber can be seen as a perturbation of a homogeneous defocusing fiber (see Fig.~\ref{fig:sketch}). We recall the latter is known to be modulationally stable. The MI gains of such random fibers with $Z_n$ as in Eq.~\eqref{eq:RWstepdistribution} and Eq.~\eqref{eq:RWPoissonstep} are illustrated in Fig.~\ref{fig:lambdabar0_RW} (for the random walk model) and in Fig.~\ref{fig:lambdabar0_Poisson} (for the Poisson model) for various parameter values $\varepsilon$ and $\epsilonZ$ as indicated.
It is worth reminding that the sample MI gain is deterministic if calculated at infinite $z$. In the numerics we calculated an approximation of the sample MI gain from Eq. (\ref{eq:kickgain}) for $n=250$ kicks and averaged over 500 realizations (dashed blue curves in the figures).
In order to check that the sample MI gain correctly predicts the growth rate of the perturbations we computed the latter from a numerical solution of the NLSE using the same procedure (black stars in the figures). One observes a good agreement.

One notices that the random perturbation produces modulational instabilities that we now further analyze, for a given distribution of the $\Delta Z_n$ and the $\lambda_n$,  and at fixed $\omega$. This can be done perturbatively in $\varepsilon$. 
Indeed, since $\overline\lambda=0$, we have $\lambda=\varepsilon\delta\lambda$, which is small. We can therefore compute the largest eigenvalue of $\overline M$ perturbatively in $\varepsilon$; its logarithm will yield the mean MI gain $G_2(\omega)$. The details of the computation are given in Appendix~\ref{appendix}.

To set up the perturbation problem, we proceed as follows. 
For $\varepsilon=0$, $\overline M=\overline M_{+}$ so that the first ingredient we need are the eigenvalues and eigenvectors of $\overline M_+$.  It is easy to check that $\overline M_+$ has the eigenvalue $x_0=1$.
The corresponding right and left eigenvectors are
\begin{equation}\label{eq:oneeigenvectors}
  \varphi^{(0)} =
  \begin{pmatrix}
    \mu^2 & 0 & 1
  \end{pmatrix}^T
  \;
  \text{and}
  \;
  \psi^{(0)} = 
  \begin{pmatrix}
    1 & 0 &  \mu^2
  \end{pmatrix}^T,  
\end{equation}
with ${\psi^{(0)}}^T \varphi^{(0)}=2\mu^2$.

We always suppose $\omega\neq 0$ so that $\mu\neq 0$. 
The two remaining eigenvalues of $\overline M_+$ are easily determined to be
\begin{equation}\label{eq:xzeropm}
\begin{aligned}
x_\pm
&=\overline{\exp(\pm i2k\Delta Z)}=\int_0^{+\infty} \exp(\pm i2k\Zref j)\rho(j)\rd j \\
&=\sqrt{2\pi}\hat \rho (\mp  2k\Zref).
\end{aligned}
\end{equation}


Since the $\exp(i2k\Zref j)$ lie on the unit circle, clearly $|x_\pm|\leq 1$. To apply nondegenerate perturbation theory, we need that $x_\pm\not=1$. To investigate this condition in the two models that we investigate here for the $Z_n$, let us first consider
the random walk model Eq.~\eqref{eq:RWstepdistribution}, and assume the distribution of $\delta\! j$ is given by a density $\sigma$, so that
\begin{equation}\label{eq:xpm}
x_\pm=\overline{\exp(\pm i2k\Delta Z)}=e^{\pm i2k\Zref}\int e^{\pm2ik\Zref\epsilonZ s}\sigma(s)\rd s.
\end{equation}
We can then compute $x_\pm$ perturbatively in $\epsilonZ$ as follows. Noting that,
when $\epsilonZ=0$, $x_\pm=\exp(\pm i2k\Zref)$ and  remembering that $\langle \delta\! j\rangle=0$, one sees
\begin{equation}\label{eq:xpm_pert}
x_\pm=\exp(\pm i2k\Zref)\left[1-2(k\Zref\epsilonZ)^2\langle \delta\! j^2\rangle+ \text{O}(\epsilonZ^3)\right],
\end{equation}
so that, with increasing $\epsilonZ$,  the eigenvalues $x_\pm$ move radially inward, towards the origin. Hence $|x_\pm|<1$ and, a fortiori, $x_\pm\not=x_0=1$ for $\epsilonZ\not=0$. 
For a Poisson fiber, where the $Z_n$ form a Poisson process, the eigenvalues $x_\pm$ can be computed explicitly:
\begin{equation}\label{eq:xpmPoisson}
x_\pm=\frac{1}{1\mp i2k\Zref},
\end{equation}
so that $|x_\pm|<1$ for all $\omega\not=0$. 

We can therefore use, in the above cases, a non-degenerate  perturbation expansion to compute $x_0$ as a function of $\varepsilon$. We will establish that $x_0$ is real and a growing function of $\varepsilon$, giving rise to a strictly positive mean MI gain. 


Recalling that $\lambda=\varepsilon \delta\lambda$ (See Eq.~\eqref{eq:lambdan}, with $\overline\lambda=0$),   we consider the case where the probability distribution $\nu(\delta\lambda)$ of $\delta\lambda$ satisfies $\nu(-\delta\lambda)=\nu(\delta\lambda)$ so that not only $\overline\lambda=\varepsilon\int s\nu(s)\rd s=0$ but also
$
\overline{\sin(2\theta)}=0,
$
since $\sin\theta$ is an odd function of $\delta\lambda$.  Hence, we can write
\begin{equation}\label{eq:overlineMlambda}
\overline{M}=\overline M_+ + \Delta\overline M\eta,
\end{equation}
where
$
\eta=\overline{\cos(2\theta)}-1=\sqrt{2\pi}\left(\hat\nu(\varepsilon \Delta\beta_2\Zref\omega^2)-\hat\nu(0)\right)$ and $\Delta\overline M=\frac12(\overline M_+-\overline M_-).$

 Finally, for small $\varepsilon\omega^2$,
\begin{equation}
\eta\simeq-\frac12(\varepsilon \Delta\beta_2\Zref\omega^2)^2\overline{\delta\lambda^2}.
\end{equation}
The eigenvalue $x_\eta$ of $\overline M$ in~ Eq. \eqref{eq:overlineMlambda} that emanates from $x_0=1$ can be expanded as
\begin{equation}
\label{eq:TaylorEV}
x_\eta \approx 1+\eta x^{(1)}+\eta^2 x^{(2)},
\end{equation}
where
\begin{align}
\nonumber   x^{(1)} &=-\frac{(1-\mu^2)^2}{4\mu^2},\\
x^{(2)}&=\frac1{16\mu^4}(1-\mu^4)^2\frac{2S(2S-1)+S_2^2}{4S^2+S_2^2},
\end{align}
with
\begin{equation}
S=\overline{\sin^2(k\Delta Z)}=\frac12\left(1-\overline{\cos(2k\Delta Z)}\right),\; S_2=\overline{\sin(2k\Delta Z)}.
\end{equation}
The corresponding mean MI gain is then, using Eq.~\eqref{eq:meangain}
\begin{equation}\label{eq:meangaineta}
G_2(\omega)=\frac{1}{2\Zref}\ln x_\eta\approx\frac1{2\Zref}\ln(1+\eta x^{(1)}+\eta^2 x^{(2)}).
\end{equation}
To lowest order in $\varepsilon$, we therefore find that, approximately, for small $\omega$, 
\begin{equation}\label{eq:firstordergainlambdabarzero}
G_2(\omega)\approx \frac12 |\eta|\frac{(1-\mu^2)^2}{4\mu^2}\approx\frac14\frac{\Delta\beta_2^2\Zref^2\gamma P}{\betaref}(\varepsilon\omega)^2\overline{\delta\lambda^2}.
\end{equation}

The mean MI gain $G_2(\omega)$, computed by evaluating the spectral radius of $\overline M$ numerically, as well as its approximation from Eq.~\eqref{eq:meangaineta},  are shown in Fig.~\ref{fig:lambdabar0_RW} for the simple random walk model and in Fig.~\ref{fig:lambdabar0_Poisson}  for the Poisson model.  There is, in both cases, a MI side lobe. 
The perturbative approximation works well for the Poisson model, for all $\omega$ in the range considered (see dashed black curves in Fig.~\ref{fig:lambdabar0_Poisson} ). It does however not capture the vanishing of the MI that occurs at a specific value of $\omega$ for $\epsilonZ=0$ or small in the random walk model (see Fig.~\ref{fig:lambdabar0_RW}). There is indeed a marked difference between the shape of these side lobes, depending on which of  the two random processes are chosen for the $Z_n$, that we now explain.

For that purpose, first consider  the leftmost column of Fig.~\ref{fig:lambdabar0_RW}. There $\epsilonZ = 0$, which means the $Z_n=n\Zref$ are distributed periodically along the fiber. The GVD is nevertheless not periodic, since the kick strengths $\lambda_n$ are random. One notices on the figure that, in that case, both the sample MI gain and the mean MI gain show a characteristic zero at a precise value of $\omega$. This phenomenon can be explained as follows. If $\omega=\omega_\ell$ is such that $k\Zref=\pi\ell$ for some $\ell\in\Z$, then  the matrices $L_n$ in Eq.~\eqref{eq:transportmatrix} are all equal to the identity matrix, and one immediately sees from Eq.~\eqref{eq:kickgain}  that $G(\omega_\ell)=0$, for all $\varepsilon$. 
 Indeed the propagation through the constant dispersion segments of the fiber does not change the perturbation and the kicks act  as random rotations, which only change the phase of the perturbation $\hat v$, giving as a results a vanishing MI gain.
Using Eq.~\eqref{eq:k2omega}, one sees this corresponds to the specific values of $\omega$ given by $\omega=\omega_\ell$, where
\begin{equation}
\omega_\ell^2=\frac{2}{\betaref}\left(\sqrt{\left(\gamma P\right)^2+\left(\frac{\pi\ell}{\Zref}\right)^2}-\gamma P \right).\label{eq:omega_ell}
\end{equation}  
One furthermore readily checks that the eigenvalues of $\overline M$ in this case are given by $1, \overline{\cos\theta}, \overline{\sin\theta}$, which are less than one in absolute value.  Therefore the mean MI gain $G_2(\omega_\ell)$ also vanishes for all $\varepsilon$ at $\omega=\omega_\ell$.  Again, this is apparent from Fig.~\ref{fig:lambdabar0_RW} at $\omega=\omega_1$.
Interestingly enough, the frequencies determined from Eq.~\eqref{eq:omega_ell}, fulfill the parametric resonance condition $k\Zref=\pi\ell$ \cite{Armaroli2012,RNCDKMTDB,Conforti2016}. These frequencies correspond to the location of the tips of Arnold tongues for any periodic fiber with period $\Zref$ and whose average GVD over one period equals $\beta_{2,\textrm{ref}}$. 
%

Quite surprisingly, for kicks with random amplitudes and zero-mean, the situation is reversed and the system becomes stable under perturbations precisely at these same frequencies.

It is finally clear from Fig.~\ref{fig:lambdabar0_RW} that the perturbative treatment of the mean MI gain does not function well when $\omega$ approaches $\omega_1$ or $\omega_2$. This is as expected since, when $\omega=\omega_\ell$, the three eigenvalues of $\overline M_+$ coincide: $x_\pm=1=x_0$. The nondegenerate perturbation theory used above does then not apply.  When $\omega\not=\omega_\ell$, this degeneracy is lifted and the perturbation theory yields increasingly good results as $\epsilonZ$ increases, even at $\omega=\omega_\ell$. As can be seen in the second column of Fig.~\ref{fig:lambdabar0_RW}, both the sample and mean MI gains  are still diminished in the neighbourhood of $\omega_1$ and $\omega_2$ when $\epsilonZ$ is small, and this is well captured by the perturbative analysis above. This phenomenon can be seen as a remnant of the underlying periodic structure of the random points $Z_n$ that is only partially destroyed when $\epsilonZ$ is nonzero, but small. It completely disappears when $\epsilonZ$ approaches its maximal possible value, which is $1$, as can be seen in the third column of the figure.

Since for the homogeneous fiber ($\beta_2(z)=\beta_{2,\textrm{ref}}$, $\epsilonZ=0=\varepsilon$) there is no MI at all, it is clear that all MI is created by the randomness. Note however that, whereas increased fluctuations in the kicking strengths $\lambda_n$ increases the MI, increased fluctuations in the $Z_n$ tends to decrease it.  

In Fig.~\ref{fig:lambdabar0_Poisson} the MI for the Poisson model is illustrated.  One sees that, as for the simple random walk model, there is an MI side lobe starting at low frequencies, 
but the frequencies $\omega_\ell$ do now no longer play a special role.  The gain of this side lobe is comparable, in width and height, for the same value $\varepsilon=0.5$ of the strength of the kicks, as in the random walk fiber with $\epsilonZ=0.9$. Note that, in the latter, this means successive $Z_n$ can be close, as in the Poisson fiber. The perturbative treatment of the previous section reproduces the mean MI gain quite accurately. 





%
%

\begin{figure*}
  \begin{center}
    \begin{tabular}{ccc}
      \includegraphics[width=0.25 \textwidth,keepaspectratio]{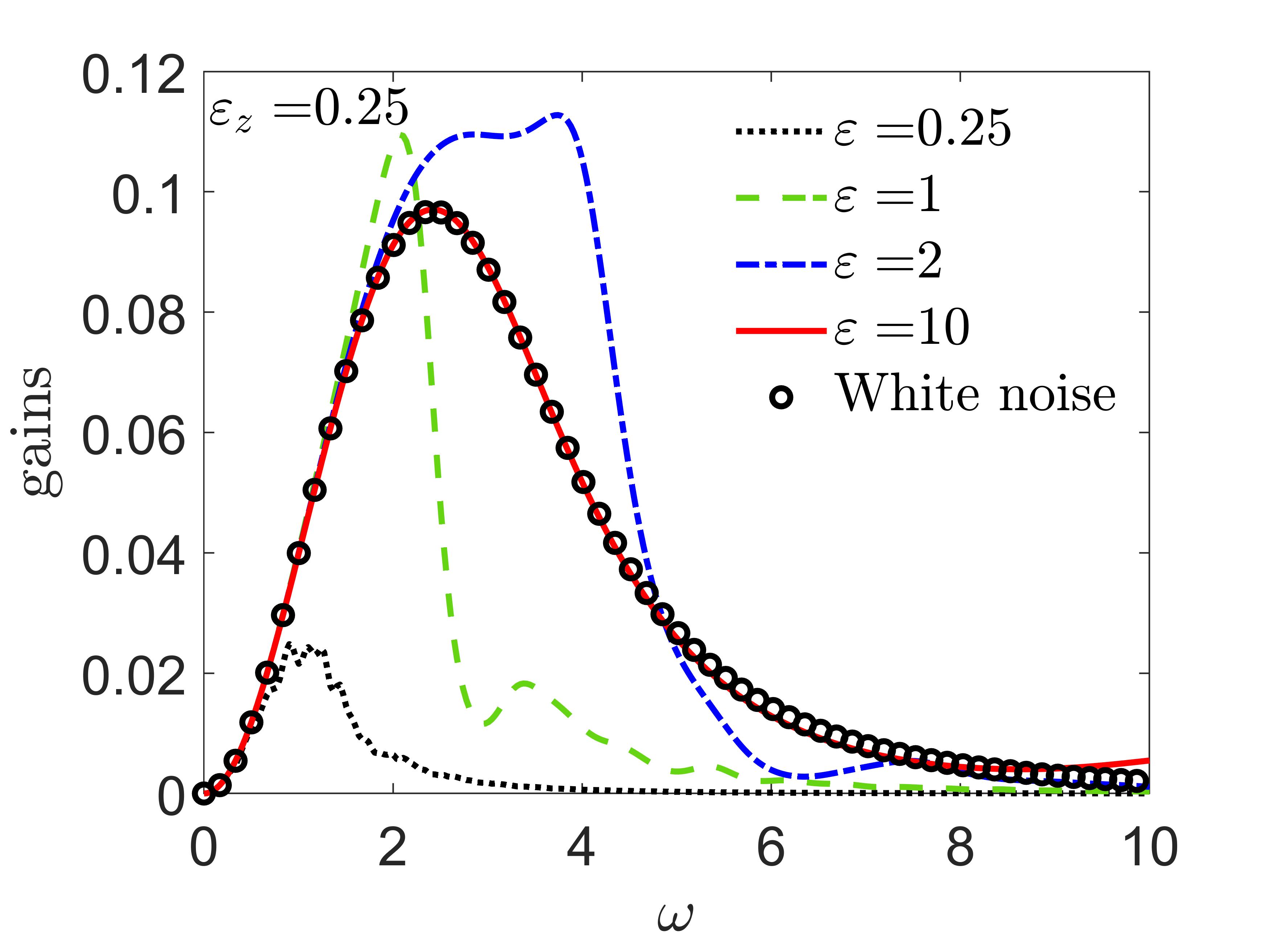} 
        \includegraphics[width=0.25 \textwidth,keepaspectratio]{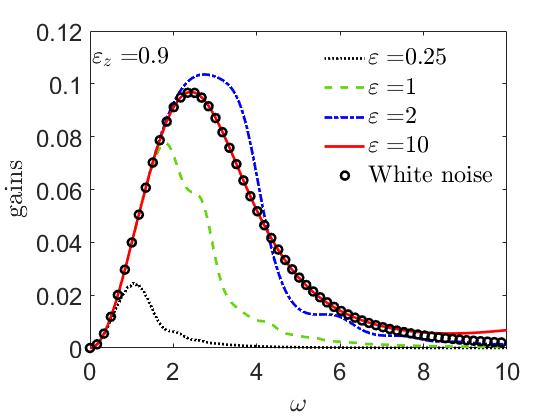}
        \includegraphics[width=0.25 \textwidth,keepaspectratio]{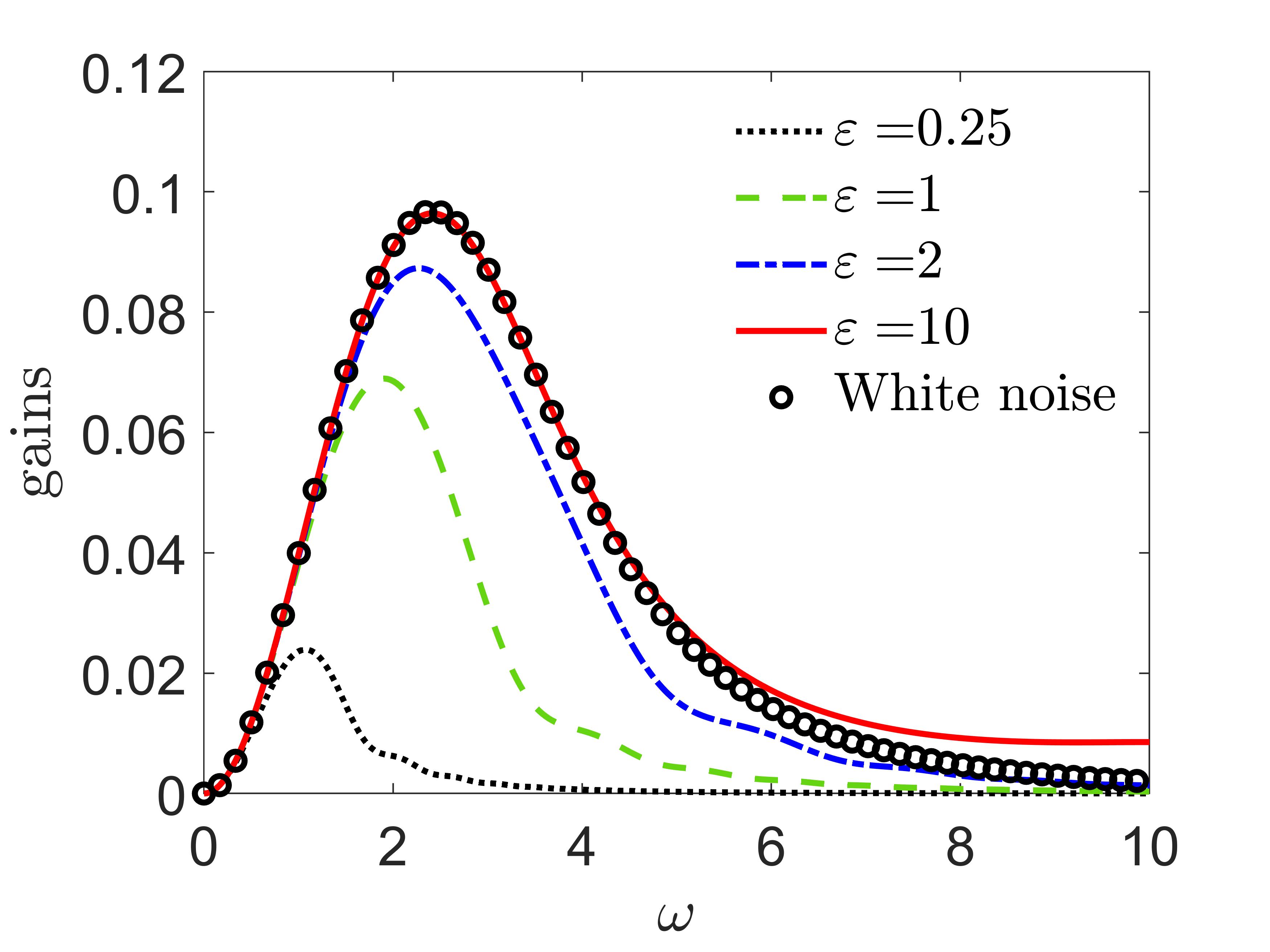} 
    \end{tabular}
  \end{center}
  \caption{Comparison between the mean MI gain of the white noise model (black dots) and the mean MI gain of the random walk (left and center panels) and Poisson models (right panel). MI gain for white noise model  is calculated from Eq. (29) form \cite{Abdullaev1996} with $A=1$  ($A\equiv\sqrt{P}$ in our notation) and $\sigma^2=0.1$. Parameters $\varepsilon$ and $\epsilonZ$  as indicated and $\Zref$ as in~\eqref{eq:ZrefWN} with $\overline{\delta\lambda^2}=1/3$.}\label{fig:WNlimit}
\end{figure*}

\section{Approximating white noise GVD with randomly kicked GVD}
Previous work on MI in random fibers has concentrated on GVD perturbed by white noise~\cite{Abdullaev1997,Abdullaev1997, Abdullaev1999, Garnier2000}. As pointed out above, white noise is not necessarily physically pertinent since it requires arbitrarily large variations of the GVD over arbitrarily short distances. 
In this section we will show that the kicked fibers considered here can, in an appropriate parameter regime determined below, and for sufficiently low frequencies, produce a similar MI gain as a white noise GVD. Fig.~\ref{fig:WNlimit} illustrates our findings. 

For a kicked GVD as in~\eqref{eq:beta2approximate}, with $\overline\lambda=0$, the two-point function is 
\begin{eqnarray*}
\overline{\left(\beta_2(z)-\beta_{2,\textrm{ref}})(\beta_2(z')-\beta_{2,\textrm{ref}}\right)}&=&\ \\
&\ &\hskip-3cm (\Delta\beta_2 Z_{\text{ref}}\varepsilon )^2\overline{\delta\lambda^2}\left(\sum_n \rho_n(z)\right)\delta(z-z').
\end{eqnarray*}
 Here the average is taken with respect to $\lambda_n$ and $Z_n$ and $\rho_n(z)$ is the probability distribution function of $Z_n$. Note that the fiber is therefore delta-correlated in $z$, but it is not stationary, since $\sum_n \rho_n(z)$ is not constant. Nevertheless, one finds
$$
\lim_{z\to+\infty}\sum_n\rho_n(z)=\Zref^{-1},
$$
so that it does becomes stationary for large $z$. To show this, one can proceed as follows. Introducing the counting function 
$
N(Z)=\sharp\{n\mid Z_n\leq Z\},
$
one easily sees that $\overline N(Z)\simeq \frac{Z}{\Zref}$. On the other hand,
$
N(Z)=\int_0^Z n(z) \rd z$, where  $n(z)=\sum_m \delta(z-Z_m).
$
So, since
$$
n(z)=\frac{\rd N}{\rd z}(z),\quad \overline{n}(z)=\frac{\rd \overline N}{\rd z}(z),
$$
it follows from 
$
\overline n(z)=\sum_m\rho_m(z)
$
that
$$
\lim_{z\to+\infty}\sum_n\rho_n(z)=\lim_{z\to+\infty}\overline{n}(z)=\frac{\rd \overline N}{\rd z}(z)=\frac1{\Zref}.
$$
In conclusion, for $z$ large, one has 
$$
\overline{\left(\beta_2(z)-\beta_{2,\textrm{ref}})(\beta_2(z')-\beta_{2,\textrm{ref}}\right)}\approx 2\sigma^2\delta(z-z'),
$$
where
\begin{equation}
2\sigma^2=(\Delta\beta_2\varepsilon)^2 Z_{\text{ref}}\overline{\delta\lambda^2}.
\end{equation}
One recognizes here the two-point function of a Gaussian white noise. In a kicked fiber, the fluctuations take place on a length scale comparable to $\Zref$ and have a strength proportional to $\varepsilon$. This suggests that, if $\Zref$ is small and $\varepsilon$ large, with a scaling given by 
\begin{equation}\label{eq:ZrefWN}
 \Zref=2\sigma^2/(\Delta\beta_2^2 \varepsilon^2 \overline{\delta \lambda^2}) 
\end{equation}
then the kicked fiber will be statistically close to a white noise fiber and the resulting MI will therefore be similar in both fibers. This is indeed illustrated in Fig~\ref{fig:WNlimit}. The mean MI gain of a white noise fiber is plotted there (black circles) using an explicit formula for this gain obtained in~\cite{Abdullaev1996}. It is compared to the mean MI gain of randomly kicked fibers, with parameters $\varepsilon$ and $\epsilonZ$  as indicated and with $\Zref$ as in~\eqref{eq:ZrefWN}, computed from the largest eigenvalue of $\overline M$. As suggested by the above argument, for sufficiently large $\varepsilon$, the MI gain of the kicked fibers converges to the one of the white noise fiber. The agreement is best for small $\omega$, a reflection of the fact that the low frequency perturbations are less  sensitive to the rapid variations of the white noise. 

We make two further comments. First, the value of $\sigma^2=0.1$ is chosen in the numerics because it is the right order of magnitude for the fibers used in the experiments described in~\cite{RNCDKMTDB} in which both $\varepsilon$ and $\Zref$ are of order $1$.  Note that this means that the mean spacing between the kicks in these fibers is of the same order of magnitude as the nonlinear length $Z_{NL}=(\gamma P)^{-1}$. 
Second, for higher but intermediate values of $\varepsilon$, the MI lobe of the randomly kicked fibers can be higher and wider than the one of the white noise model. Third, it would be a challenge to make fibers with a considerably larger value of  $\varepsilon$ since they correspond to a small value of $\Zref$. This means that one would need to be able to put the sharp peaks and dips in the fiber diameter very closely together.

\begin{figure*}[t]
  \begin{center}
    \begin{tabular}{ccc}
      \includegraphics[width=0.25 \textwidth,keepaspectratio]{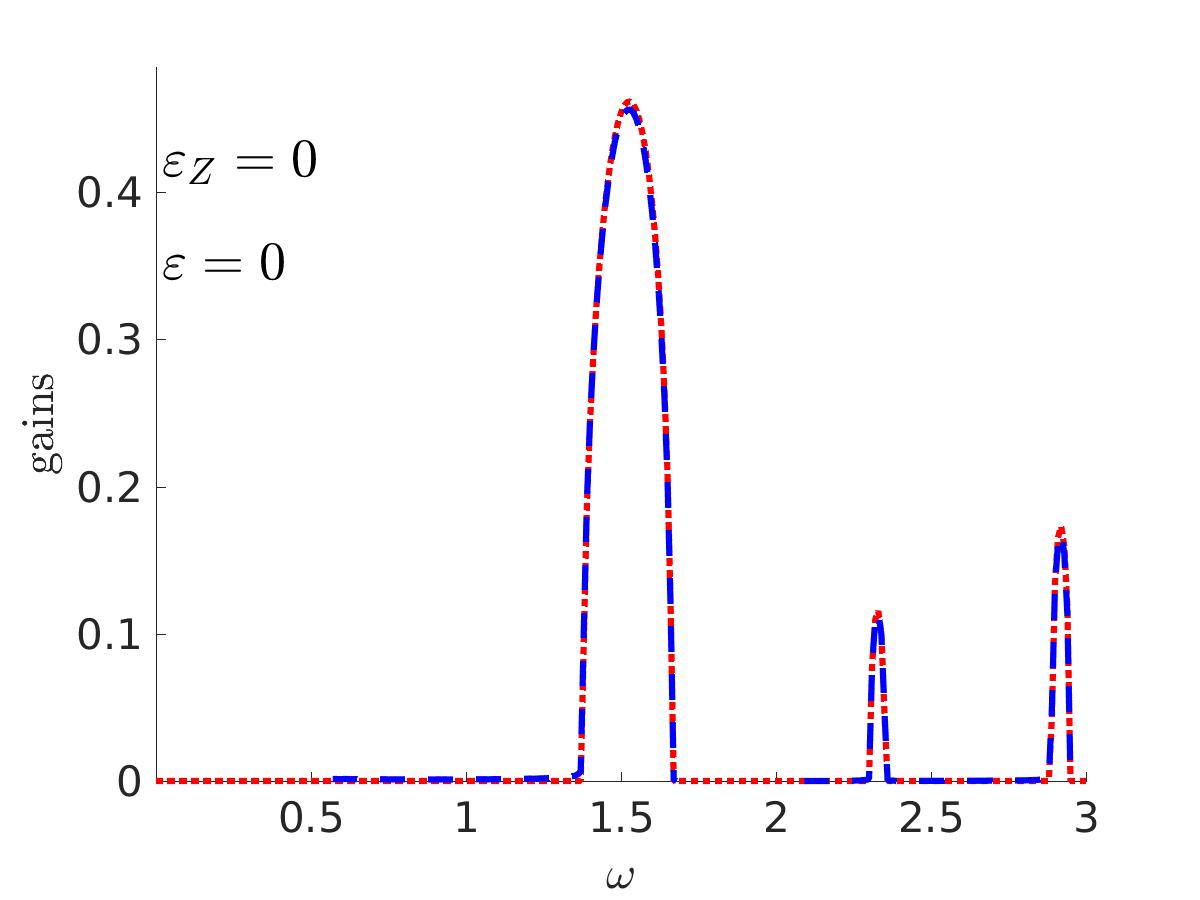}&
      \includegraphics[width=0.25 \textwidth,keepaspectratio]{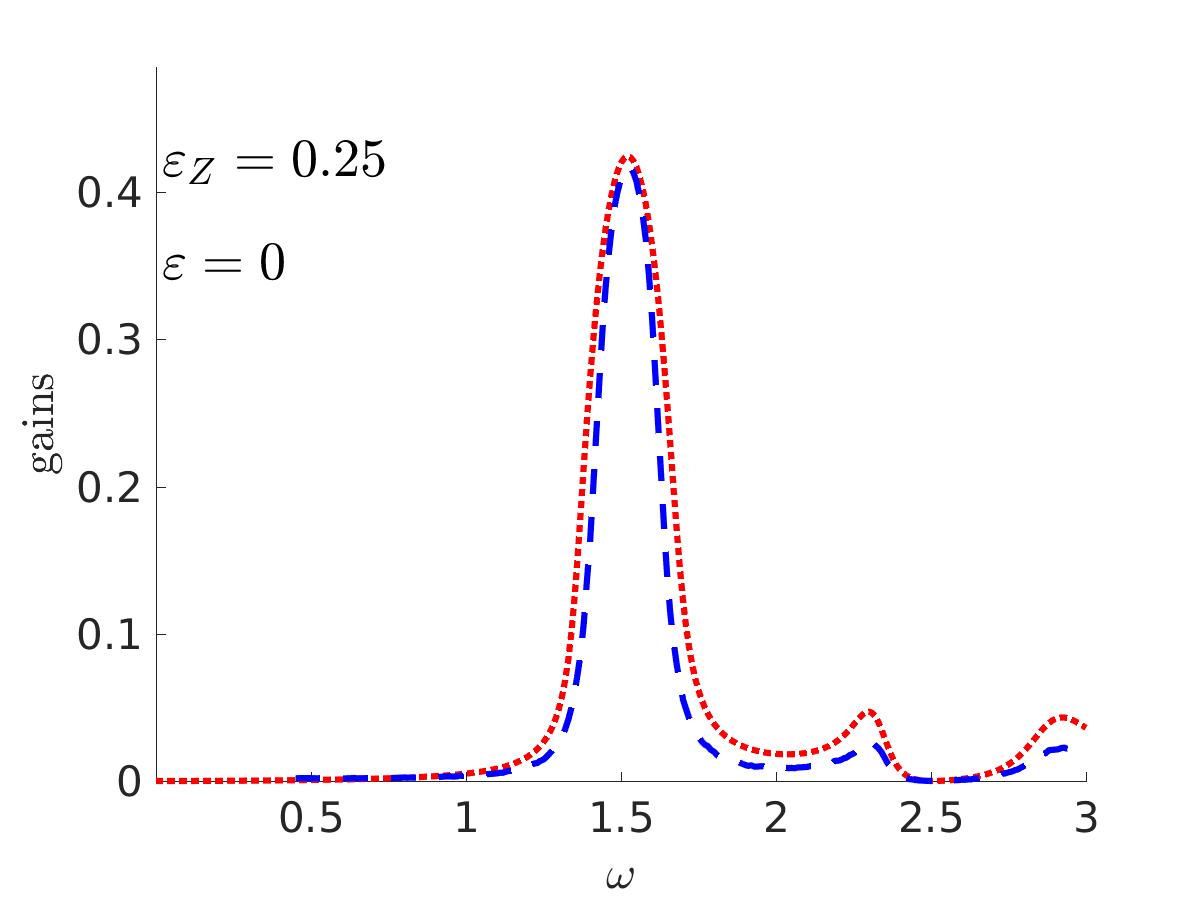}&
      \includegraphics[width=0.25 \textwidth,keepaspectratio]{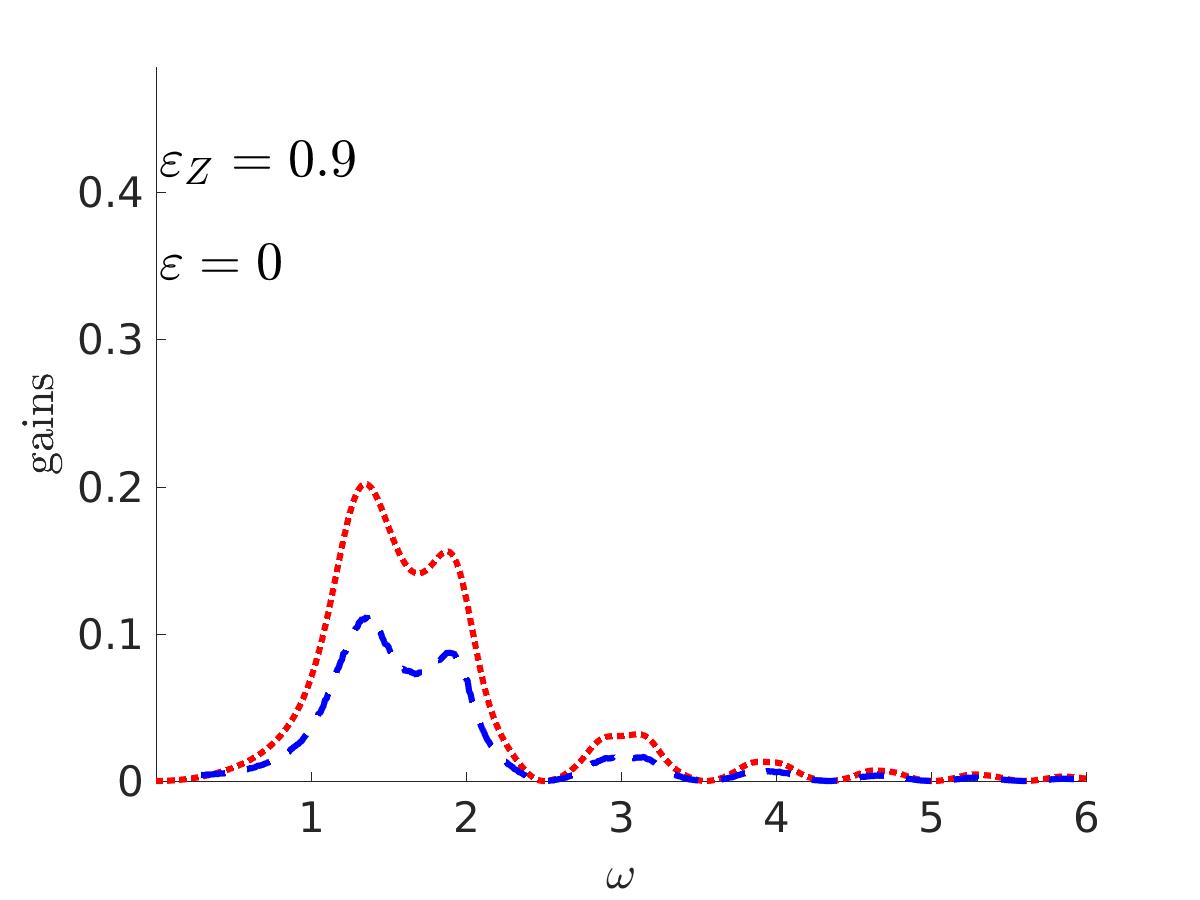}\\
      \includegraphics[width=0.25 \textwidth,keepaspectratio]{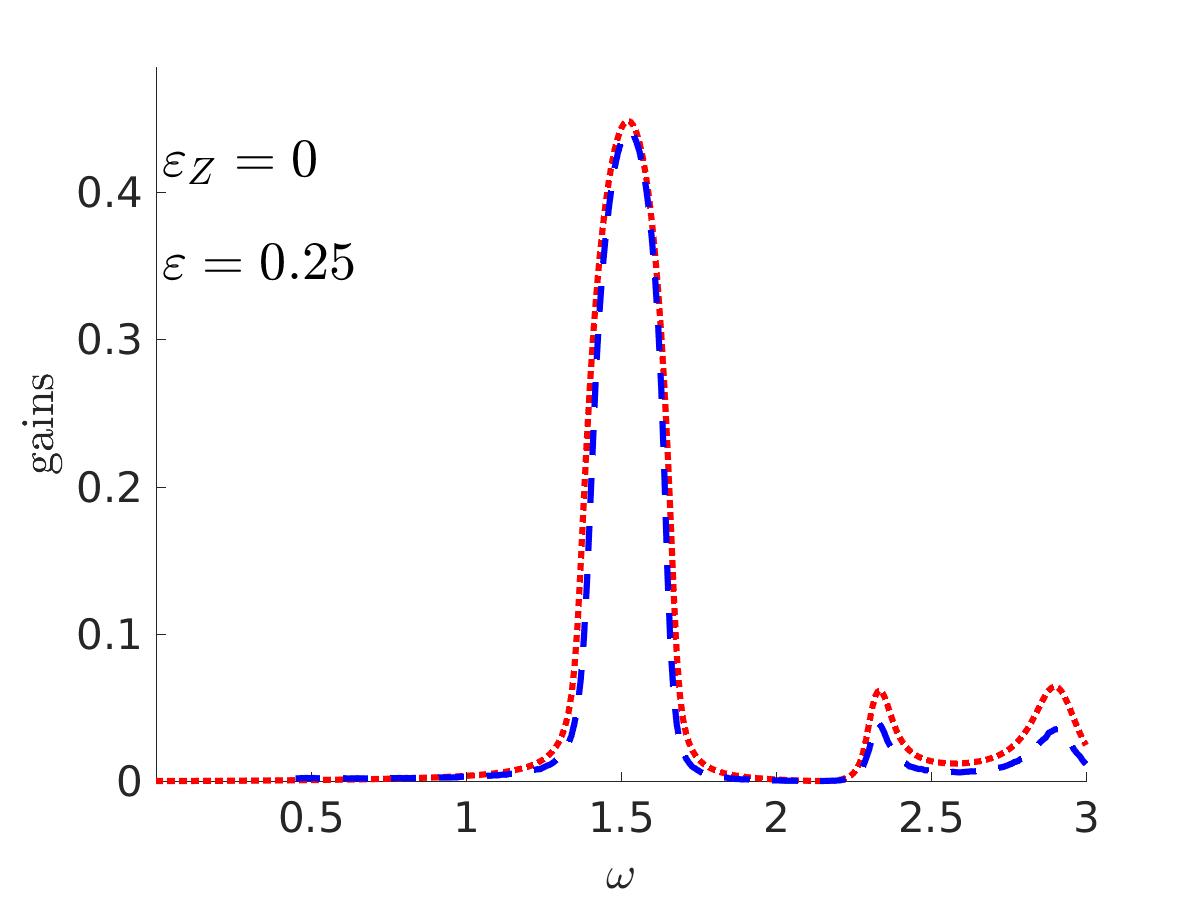}&
      \includegraphics[width=0.25 \textwidth,keepaspectratio]{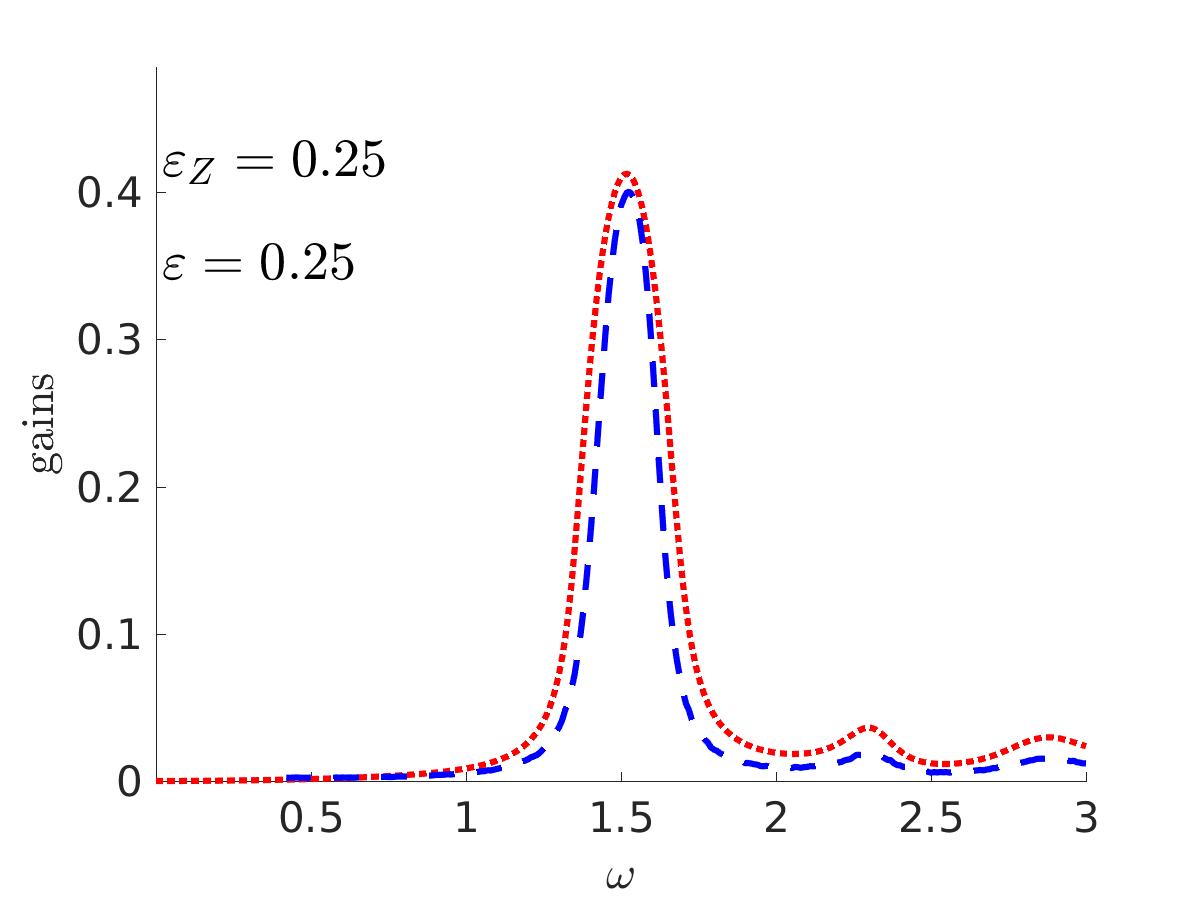}&
      \includegraphics[width=0.25 \textwidth,keepaspectratio]{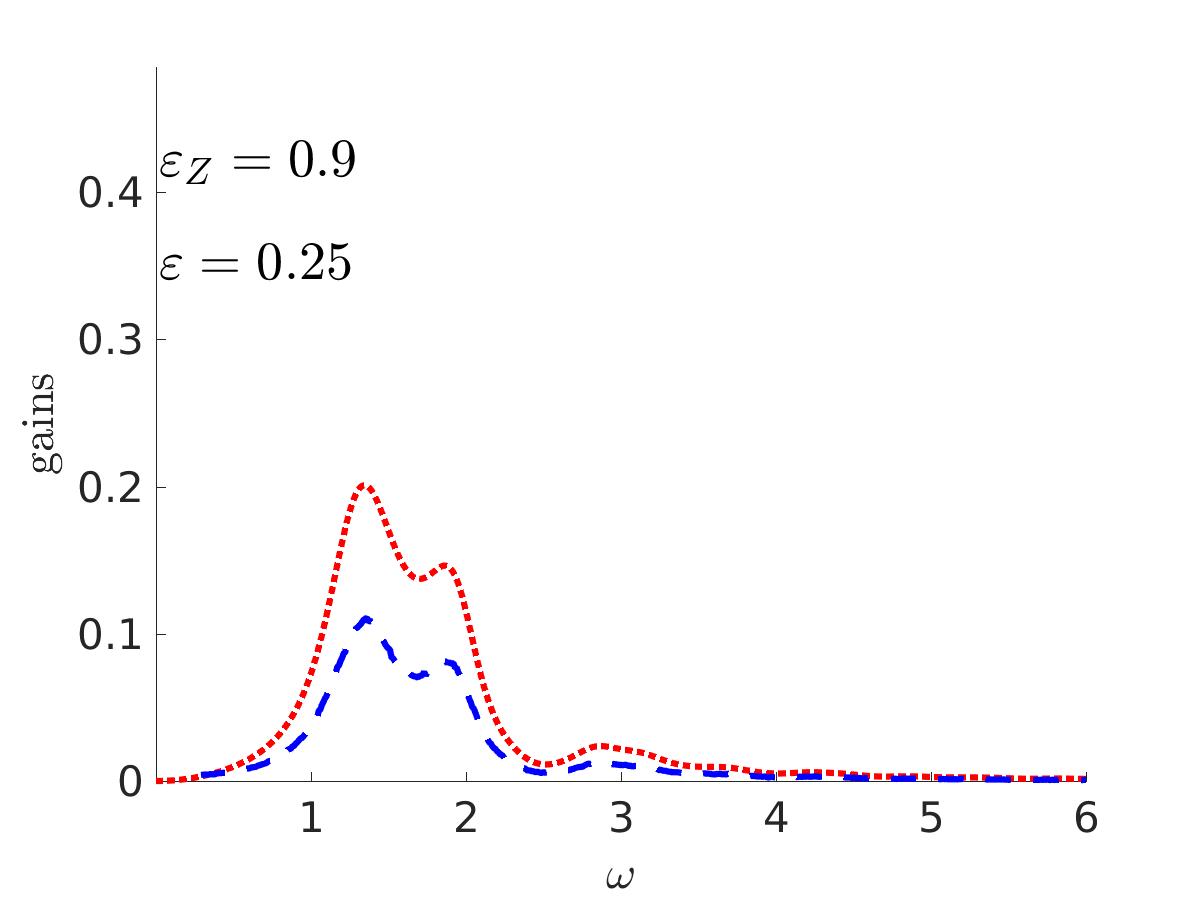}\\
      \includegraphics[width=0.25 \textwidth,keepaspectratio]{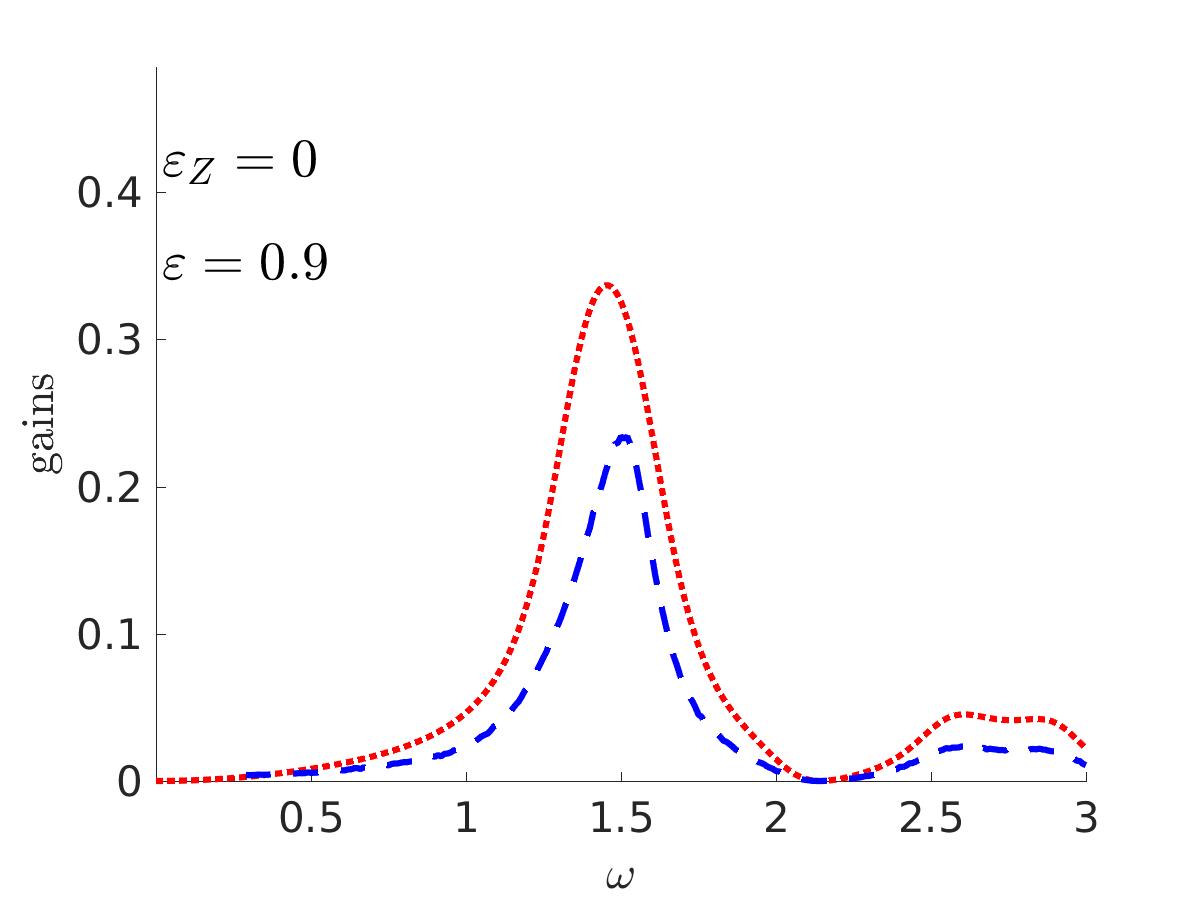}&
      \includegraphics[width=0.25 \textwidth,keepaspectratio]{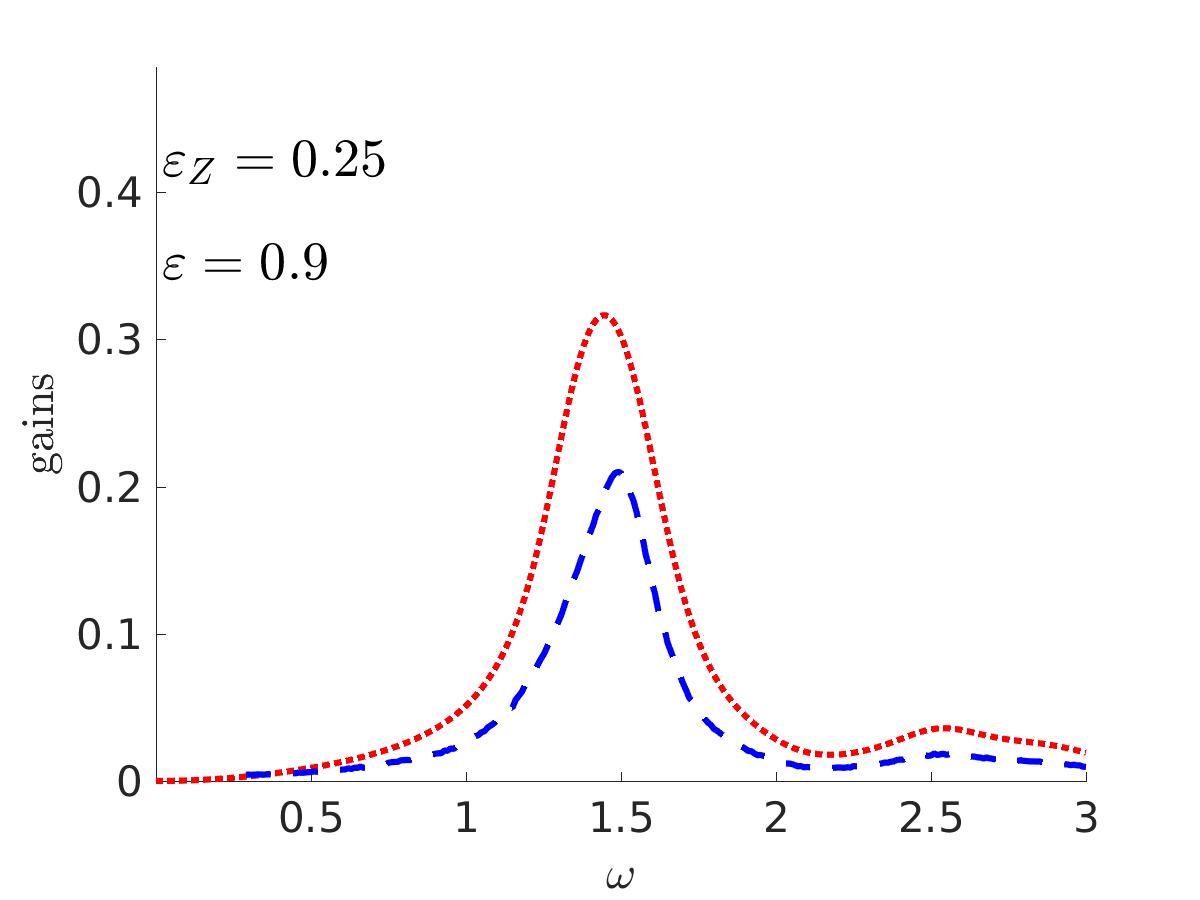}&
      \includegraphics[width=0.25 \textwidth,keepaspectratio]{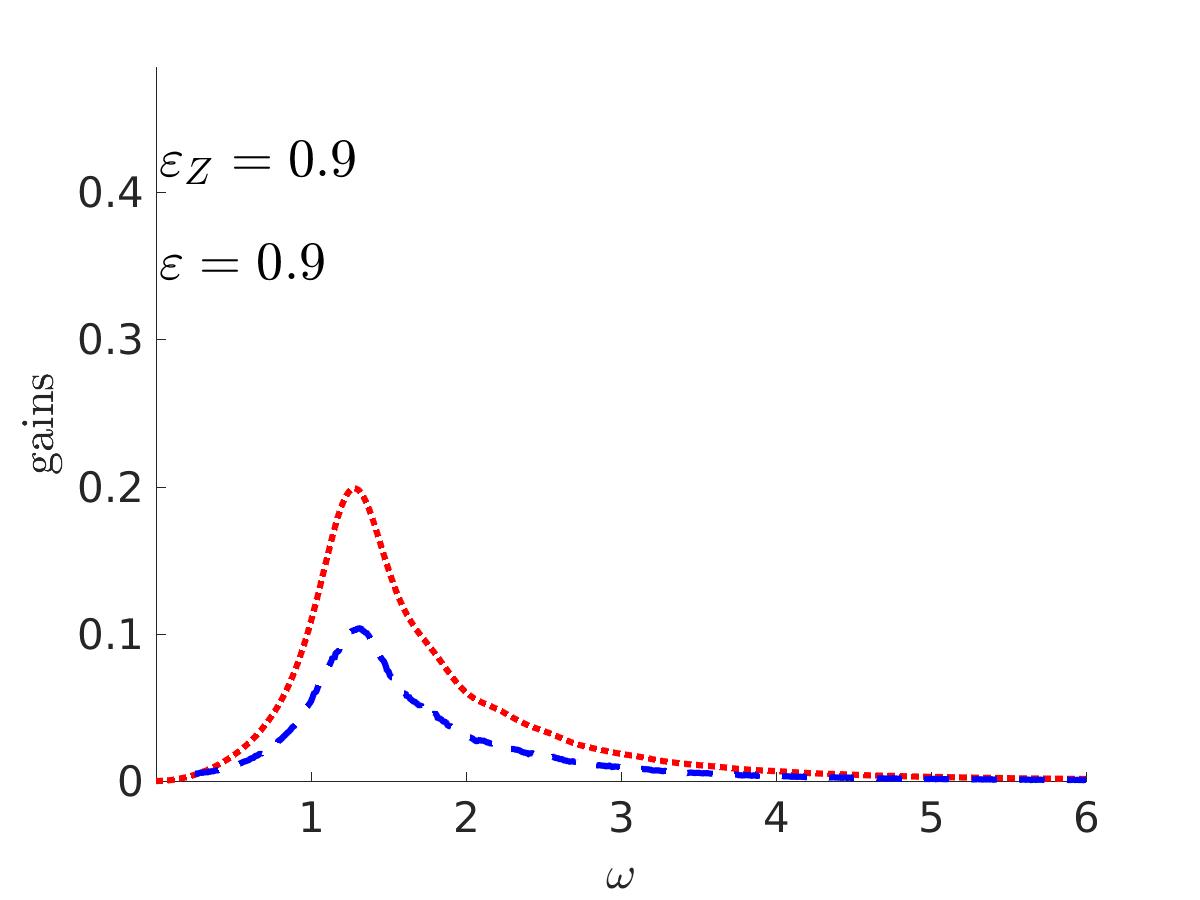}
    \end{tabular}
  \end{center}
  \caption{Sample and mean MI gains for a simple random walk process and $\overline\lambda=1$ and $\delta\lambda_n$ uniform in $[-1,1]$. 
  Color code is the same as Fig.~\ref{fig:lambdabar0_RW}} \label{fig:lambdabar1_RW}
\end{figure*}

\section{Nonzero average kick amplitude}\label{s:defoclambdaplus}

In this section we briefly discuss the effect on MI of randomly placed kicks with  random strengths $\lambda_n$ that are not vanishing on average, so that $\overline \lambda\not=0$. We will separately consider the two cases where the $Z_n$ form a simple random walk or a Poisson process, and that lead to different phenomena. 

We first consider the simple random walk, as in Eq.~\eqref{eq:RWstepdistribution}, an example of which is reported in Fig. \ref{fig:sketch}, rightmost panel. In the first column of Fig.~\ref{fig:lambdabar1_RW} the numerically computed sample and mean MI  gains are displayed for such a fiber for which $\epsilonZ=0$, so that $Z_n=n\Zref$, \textit{i.e.}, the delta kicks are placed periodically.   The $\lambda_n$ are chosen as in Eq.~\eqref{eq:lambdan} with average $\overline\lambda=1$. Note that, if in addition $\varepsilon=0$, then this fiber is actually periodic: this is the situation in the top left panel of the figure. It is well known that periodic fibers display MI through so-called Arnold tongues \cite{LANDAU197658,ArnoldMMCM,Nayfeh1979,Armaroli2012,Armaroli2013} and this has been experimentally shown in periodically kicked fibers~\cite{RNCDKMTDB,Mussot2018}.  The first such Arnold tongue can be observed in the top left panel of Fig.~\ref{fig:lambdabar1_RW}. In the other panels of the first column, $\varepsilon\not=0$ so that the kick strengths are now random, while the $Z_n$ remain periodic.  The corresponding fibers can therefore be viewed as random perturbations of a periodic fiber. It appears from these data that the fluctuating kick strengths only weakly affect the position and strength of the Arnold tongues. In the other columns of Fig.~\ref{fig:lambdabar1_RW}, on the other hand, MI gains are shown for fibers for which there is randomness in \emph{both} kick positions $Z_n$ and kick strengths $\lambda_n$. It appears from these data that randomizing the positions of the kicks has a much stronger effect than randomizing their strengths. As the randomness increases, one observes a marked decrease in the MI gain, with a widening of the Arnold tongue, while its position is less affected. Note that the MI is considerably larger in these fibers with $\overline\lambda=1$ than with those having $\overline\lambda=0$ as can be seen by comparing the vertical scale of Fig.~\ref{fig:lambdabar1_RW} with that of Fig.~\ref{fig:lambdabar0_RW} and Fig.~\ref{fig:lambdabar0_Poisson}. In the first case, we are dealing with random perturbations of a periodic fiber, which displays MI due to parametric resonance. In the second, the MI is on the contrary generated by the randomness.

\begin{figure*}
  \begin{center}
    \begin{tabular}{ccc}
      \includegraphics[width=0.25 \textwidth,keepaspectratio]{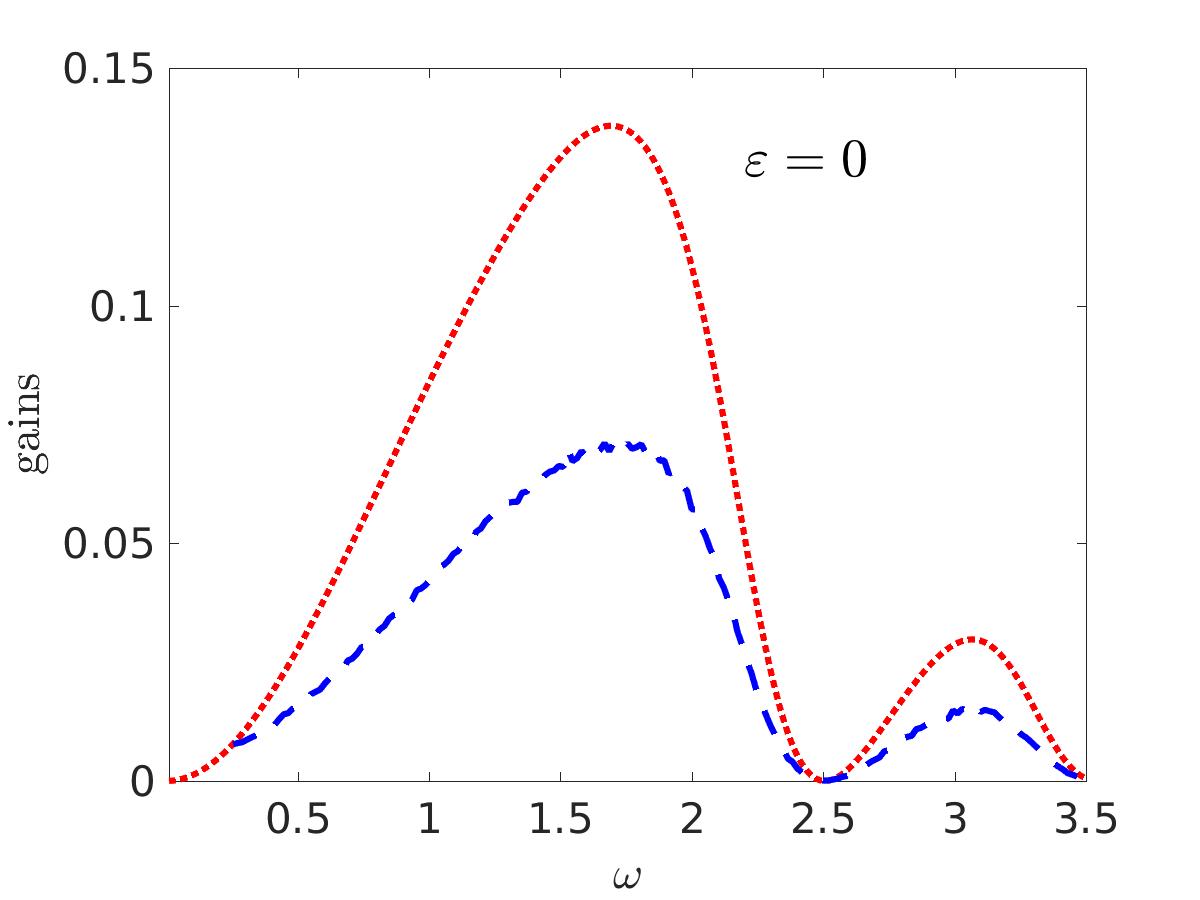}
      \includegraphics[width=0.25 \textwidth,keepaspectratio]{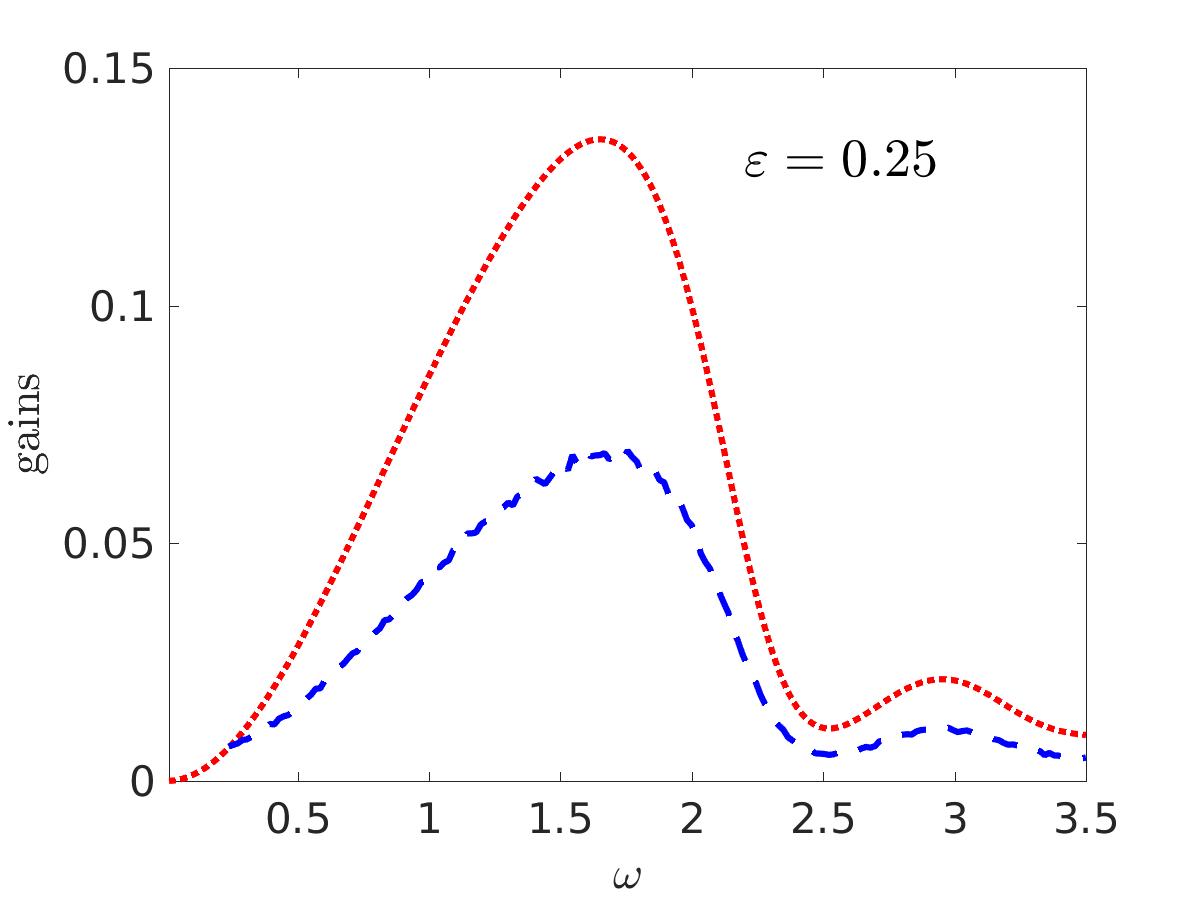}
      \includegraphics[width=0.25 \textwidth,keepaspectratio]{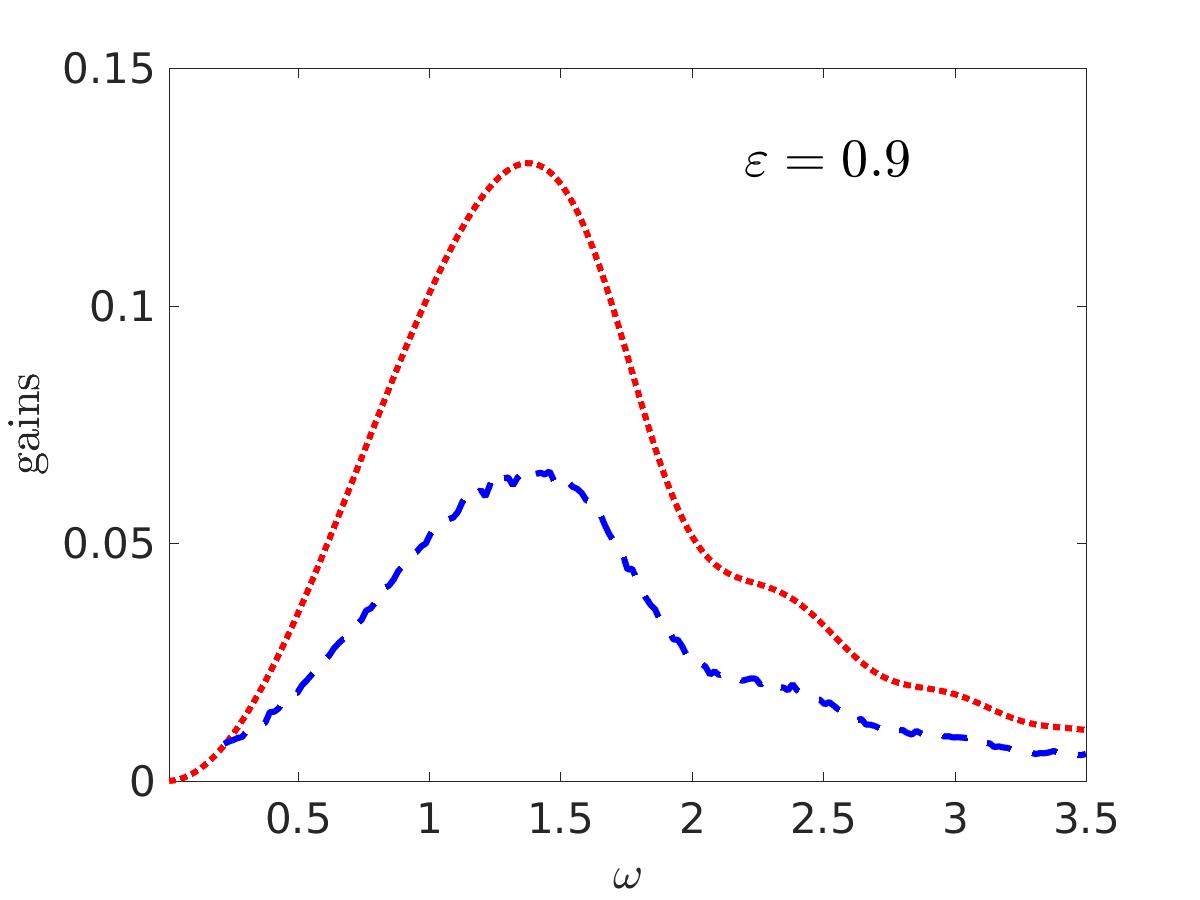}
    \end{tabular}
  \end{center}
  \caption{ Sample and mean MI gains for a Poisson model with $\overline\lambda=1$. 
   Color code is the same as Fig.~\ref{fig:lambdabar0_RW} }\label{fig:lambdabar1_RWPoisson}
\end{figure*}

In Fig.~\ref{fig:lambdabar1_RWPoisson} the numerically computed sample and mean MI gains are displayed for a Poisson fiber with values of $\varepsilon$ as indicated. For $\varepsilon=0$ one observes the identical vanishing of these gains at $\omega\approx 2.5$. 
This is actually a more general phenomenon, not related to the distribution of the kicks, that is easily understood as follows. Let us consider a defocusing fiber with, for all $n$, $\lambda_n=\overline \lambda>0$ and with arbitrary values for $Z_n$. They could in particular be periodic, quasiperiodic, or random.  If we now choose $\omega_n$ as 
\begin{equation}
n\pi=\overline\lambda\Delta \beta_{2}\Zref \frac{\omega_n^2}{2},\label{eq:zerogain}
\end{equation}
then we immediately see from (6) that $K_n=\pm I_2$. In other words, for these values of $\omega$, the kicks have no effect on the linearized solution of the equation of motion. One therefore has $\Phi_n=\pm L_n$ and so the sample MI gain is equal to the sample MI gain of the unperturbed fiber. Since the latter is defocusing, the sample MI gain vanishes. 
When $n=1$, one finds $\omega=\sqrt{2\pi}\approx 2.5$. That the mean MI gain must also vanish follows from observing that, if the above condition is satisfied, then $\theta=n\pi$ in Eq.~\eqref{eq:theta}. Hence Eq.~\eqref{eq:Mtheta} implies $M=M_+$ and consequently $\overline M=\overline M_+$. But we saw that the eigenvalues of $\overline M_+$ are $1,\, x_\pm$, with $|x_{\pm}|\leq 1$. Hence the mean MI gain also vanishes. 

Let us stress that this phenomenon is not limited to the Poisson fiber, for which it can be observed in Fig.~\ref{fig:lambdabar1_RWPoisson}. In fact, for the simple random walk fiber it is visible in the top panels of Fig.~\ref{fig:lambdabar1_RW}. 
The phenomenon also appears in periodically kicked fibers where
$$
\beta_2(z)=\beta_{2, \textrm{ref}}+\Delta \beta_{2} \overline \lambda \sum_n \delta\left(\frac{z-n\Zref}{\Zref}\right),
$$
so that the spatial average of the GVD is
$$
\beta_{2,\textrm{av}}=\frac1{\Zref}\int_0^{\Zref} \beta_{2}(z)\rd z= \beta_{2, \textrm{ref}}+\Delta \beta_{2} \overline\lambda.
$$
It is then well known that the Arnold tongues for such a fiber occur at values $\omega'_n$ defined by 
$$
\beta_{2,\textrm{av}}\Zref\frac{{\omega'_n}^2}{2}=\sqrt{(\gamma P\Zref)^2 +(n\pi)^2}-\gamma P\Zref.
$$
Comparing this to Eq.~\eqref{eq:zerogain}, one observes that, for all $n$, $\omega'_n<\omega_n$. Note that both $\omega_n$ and $\omega'_n$ depend on $\overline \lambda$.  

Condition (\ref{eq:zerogain}) is complementary to condition (\ref{eq:omega_ell}). The former arises when the spacing between the kicks is constant, whereas the second appears when the amplitude of the kicks is. The two conditions show that whenever either the position or the amplitude of the kicks are deterministic, particular values of the frequency exist where the gain vanishes.

\section{Discussion and conclusions}\label{s:conclusions}
We reported on the modulation instability phenomenon in optical fibers where the GVD is a random process.  Most studies of such random fibers have concentrated on the case when a homogeneous GVD is perturbed by a stationary white noise. In that case various methods exist to compute the MI gain. White noise is however very particular and not always adequate to model physically relevant scenarios.
In this paper we investigated the behavior of the MI gain in random fibers for which the GVD is of a very different nature.
More specifically, we considered a class of experimentally realizable~\cite{RNCDKMTDB} random fibers in which a constant normal GVD is perturbed by a sequence of delta kicks with random positions and amplitudes.

The main result of our analysis is fourfold. First, we show that in this situation low frequency MI lobes always arise as a result of such random perturbations, independently of the statistical distribution of the position and amplitudes of the kicks.  We trace the occurrence of MI in random fibers to a mechanism familiar from the study of Anderson localization and of the occurrence of positive Lyapounov exponents in chaotic dynamical systems. 
Second, we show that the specific shape of these side lobes does depend on these statistical properties and we provide expressions to determine them. In particular, we find that if either the positions or the amplitudes of the kicks are deterministic, then at special frequency values the MI gain is identically zero.  At these same points, the MI remains suppressed when the random fluctuations of the control parameter remain small enough. Third, we show that the randomly kicked fibers considered behave, in a suitable parameter regime that we identify, as a fiber with a white noise GVD. Finally, we have observed that for comparable parameter regimes, the MI produced through parametric resonance in fibers with a periodic GVD is considerable larger than the MI gain obtained from random perturbations of the fibers.


\vskip 1cm
\section*{Acknowledgments}
The work was supported in part the by the French government through the Programme Investissement d’Avenir with the Labex CEMPI (Grant ANR-11-LABX-0007-01) and the I-SITE ULNE ( Grant ANR-16-IDEX-0004 ULNE, projects VERIFICO, EXAT, FUNHK)  managed by the Agence Nationale de la Recherche. The work was also supported by the Nord-Pas-de-Calais Regional Council and the European Regional Development Fund through the Contrat de Projets \'Etat-R\'egion (CPER) and IRCICA.

\appendix

\section{Perturbative analysis} \label{appendix}
The eigenvalue $x_\eta$ of $\overline M$ in~ Eq. \eqref{eq:overlineMlambda} that emanates from $1$ can be expanded as
\begin{equation}
x_\eta = 1+\eta x^{(1)}+\eta^2 x^{(2)} + \mathcal{O}(\eta^3),
\end{equation}
and the corresponding eigenvectors of $\overline M$ and $\overline M^T$,
\begin{align*}
  \varphi_\eta &= \varphi^{(0)}+\eta \varphi^{(1)} + \eta^2\varphi^{(2)} + \mathcal{O}(\eta^3),\\
  \psi_\eta &= \psi^{(0)}+ \eta \psi^{(1)} + \eta^2\psi^{(2)} +\mathcal{O}(\eta^3),
\end{align*}
where $\varphi^{(0)}, \psi^{(0)}$ are defined in Eq.~\eqref{eq:oneeigenvectors}. Expanding the eigenvalue equation in powers of $\eta$ leads in the usual manner to the first correction to the eigenvalue: 
\begin{align*}
  x^{(1)} &=\frac12 \frac{{\psi^{(0)}}^T (\overline M_+-\overline M_-) \varphi^{(0)}}{{\psi^{(0)}}^T \varphi^{(0)} }\\
          &=\frac{1}{4\mu^2}\left[2\mu^2-(1+\mu^4)\right]=-\frac{(1-\mu^2)^2}{4\mu^2}.
  \end{align*}
Note that this correction does not at all depend on the distribution of the $Z_n$ nor of the $\lambda_n$.

To obtain a satisfactory expression for the mean MI gain $G_2(\omega)$, we need to obtain the second order correction $x^{(2)}$. For that purpose, we need the first correction $\varphi^{(1)}$ to the eigenvector corresponding to $x_\eta$, which is given by
\begin{equation*}
\varphi^{(1)}=\frac1{\psi_+^T\varphi_+}\left[\left(\psi_+^T\varphi^{(1)}\right)\varphi_++\left(\psi_-^T\varphi^{(1)}\right)\varphi_-\right]
\end{equation*}
where
\begin{equation}
\psi_\pm^T\varphi^{(1)}=\frac{\psi_\pm^T(\Delta\overline M)\varphi^{(0)}}{1-x_\pm},\; \psi^{(0)T}\varphi^{(1)}=0.
\end{equation}
Here $\varphi_{\pm}$ are the eigenvectors of $\overline M_+$ and of $\overline M_+^T$ corresponding to the eigenvalues $x_\pm$, given by
\begin{align}\label{eq:eigenvectors}
\nonumber \varphi_{\pm}&=
 \begin{pmatrix} 
    -\mu^2 &\pm i\mu & 1
  \end{pmatrix}^T,\\
  \psi_\pm &= 
  \begin{pmatrix}
    1 & \pm 2i\mu &  -\mu^2
  \end{pmatrix}^T, 
\end{align}
with $\psi_\pm^T\varphi_\pm = -4\mu^2$.
Finally, the second order correction to the eigenvalue is
\begin{equation}
x^{(2)}=\frac{\psi^{(0)T}\Delta\overline MR\Delta\overline M\varphi^{(0)}}{\psi^{(0)T}\varphi^{(0)}},
\end{equation}
with
\begin{equation*}
 R=\frac1{\psi_+^T\varphi_+}\left[\frac{\varphi_+\psi_+^T}{1-x_+}+\frac{\varphi_-\psi_-^T}{1-x_-}\right]=\frac2{\psi_+^T\varphi_+}\mathrm{Re}\frac{\varphi_+\psi_+^T}{1-x_+}.
\end{equation*}
Hence
\begin{align}
\nonumber x^{(2)}&=\frac2{\psi^{(0)T}\varphi^{(0)}\psi_+^T\varphi_+}\mathrm{Re}\frac{\psi^{(0)T}\Delta\overline M\varphi_+\psi_+^T\Delta\overline M\varphi^{(0)}}{1-x_+}\\
&=\frac1{2\psi^{(0)T}\varphi^{(0)}\psi_+^T\varphi_+}\mathrm{Re}\frac{\psi^{(0)T}\overline M_-\varphi_+\psi_+^T\overline M_-\varphi^{(0)}}{1-x_+}.
\end{align}
Here we used in the last line that $\psi^{(0)T}\overline M_+\varphi_+=0$ since $\psi^{(0)T}\varphi_+=0$ and similarly $\psi_+^T\overline M_+\varphi^{(0)}=0$. One finds
\begin{equation*}
\psi_0^T\overline M_-\varphi_+=(1-\mu^4),\qquad \psi_+^T\overline M_-\varphi^{(0)}=(1-\mu^4)x_+.
\end{equation*}
Hence
\begin{align}
\nonumber x^{(2)}&=\frac1{16\mu^4}(1-\mu^4)^2\left(1+\mathrm{Re}\frac1{x_+-1}\right)\\
&=\frac1{16\mu^4}(1-\mu^4)^2\frac{2S(2S-1)+S_2^2}{4S^2+S_2^2},
\end{align}
where
\begin{equation}
S=\overline{\sin^2(k\Delta Z)}=\frac12\left(1-\overline{\cos(2k\Delta Z)}\right),\; S_2=\overline{\sin(2k\Delta Z)}.
\end{equation}


\section{Computation of $\overline{M}$ \label{M}}

We describe here how to calculate the coefficients of the $3\times 3$-matrix $\overline{M}$ in Eq.~\eqref{eq:averageM}.  They depend on the following averages of trigonometric functions.
For the 3 terms involving the kicks we have
\begin{eqnarray*}
  \overline{\cos^2\theta} & = & \frac12 \int_{-1}^1\cos^2\left[\Delta\beta_2 Z_{\rm ref} \frac{\omega^2}{2} (\overline{\lambda}+\varepsilon x) \right] {\rm d} x\\
                                                                                      & = & \frac12+\frac12 \frac{\cos\left(\overline{\lambda}\Delta\beta_2 Z_{\rm ref} \omega^2\right)\sin\left(\varepsilon\Delta\beta_2 Z_{\rm ref} \omega^2\right)}{\varepsilon\Delta\beta_2 Z_{\rm ref} \omega^2},
\end{eqnarray*}

\begin{eqnarray*}
  \overline{\sin^2\theta} & = & \frac12-\frac12 \frac{\cos\left(\overline{\lambda}\Delta\beta_2 Z_{\rm ref} \omega^2\right)\sin\left(\varepsilon\Delta\beta_2 Z_{\rm ref} \omega^2\right)}{\varepsilon\Delta\beta_2 Z_{\rm ref} \omega^2},
\end{eqnarray*}
and
\begin{equation*}
  \overline{\sin 2\theta
  }
  =
   \frac{\sin\left(\overline{\lambda}\Delta\beta_2 Z_{\rm ref} \omega^2\right)\sin\left(\varepsilon\Delta\beta_2 Z_{\rm ref} \omega^2\right)}{\varepsilon\Delta\beta_2 Z_{\rm ref} \omega^2}.
\end{equation*}

For the four terms involving $\Delta Z$, we have
for the random walk model
\begin{equation*}
  \overline{\cos^2\left(k\Delta Z\right)} = \frac12+\frac12 \cos(2kZ_{\rm ref})  \frac{\sin(2kZ_{\rm ref}\varepsilon_Z)}{2kZ_{\rm ref}\varepsilon_Z},
\end{equation*}
\begin{equation*}
  \overline{\sin^2\left(k\Delta Z\right)} = \frac12-\frac12 \cos(2kZ_{\rm ref})  \frac{\sin(2kZ_{\rm ref}\varepsilon_Z)}{2kZ_{\rm ref}\varepsilon_Z},
\end{equation*}
\begin{equation*}
  \overline{\cos\left(2k\Delta Z\right)}
  = \cos(2kZ_{\rm ref}) \frac{\sin(2k\varepsilon_Z Z_{\rm ref})}{2k\varepsilon_Z Z_{\rm ref}},
\end{equation*}
and
\begin{equation*}
  \overline{\sin\left(2k\Delta Z\right)}
  = \sin(2kZ_{\rm ref}) \frac{\sin(2k\varepsilon_Z Z_{\rm ref})}{2k\varepsilon_Z Z_{\rm ref}}.
\end{equation*}

For the four terms involving $\Delta Z$, we have
for the Poisson model
\begin{equation*}
  \overline{\cos^2\left(k\Delta Z\right)} = \frac12+\frac12 \frac{1}{1+4k^2Z_{\rm ref}^2},
\end{equation*}
\begin{equation*}
  \overline{\sin^2\left(k\Delta Z\right)} = \frac12-\frac12 \frac{1}{1+4k^2Z_{\rm ref}^2},
\end{equation*}
\begin{equation*}
  \overline{\cos\left(2k\Delta Z\right)}
  = \frac{1}{1+4k^2 Z_{\rm ref}^2},
\end{equation*}
and
\begin{equation*}
  \overline{\sin\left(2k\Delta Z\right)}
  = \frac{2kZ_{\rm ref}}{1+4k^2Z_{\rm ref}^2}.
\end{equation*}


%

\end{document}